\documentclass[traditabstract]{aa}

\usepackage[english]{babel}
\usepackage{graphicx,amsmath,amssymb,gensymb}
\usepackage{txfonts}

\usepackage[colorlinks=true, linkcolor=blue, citecolor=blue, urlcolor=blue]{hyperref}
\usepackage[singlelinecheck=true,format=default,font=small,labelfont=bf]{caption}
\usepackage{color}
\usepackage{xcolor}

\usepackage{siunitx}
\newcommand{\um}[1]{\SI{#1}{\micro\meter}}

\begin{document}

\title{High-resolution, 3D radiative transfer modelling \\ IV. AGN-powered dust heating in NGC~1068}

\author{S. Viaene\inst{\ref{inst-UGent},\ref{inst-UHerts}}
\and A. Nersesian\inst{\ref{inst-Athens-Obs}, \ref{inst-Athens-Uni}, \ref{inst-UGent}}
\and J. Fritz\inst{\ref{inst-UNAM}}
\and S. Verstocken\inst{\ref{inst-UGent}}
\and M. Baes \inst{\ref{inst-UGent}}
\and S. Bianchi \inst{\ref{inst-Firenze}}
\and V. Casasola \inst{\ref{inst-Bologna}, \ref{inst-Firenze}}
\and L. Cassar\`a \inst{\ref{inst-Milano}, \ref{inst-Athens-Obs}}
\and C. Clark \inst{\ref{inst-STSCI}}
\and J. Davies \inst{\ref{inst-Cardiff}}
\and I. De Looze \inst{\ref{inst-UGent}, \ref{inst-UCL}}
\and P. De Vis \inst{\ref{inst-Cardiff}}
\and W. Dobbels \inst{\ref{inst-UGent}}
\and M. Galametz \inst{\ref{inst-Saclay}}
\and F. Galliano \inst{\ref{inst-Saclay}}
\and A. Jones \inst{\ref{inst-IAS}}
\and S. Madden \inst{\ref{inst-Saclay}}
\and A. Mosenkov \inst{\ref{inst-StPetersB-obs}, \ref{inst-StPetersB-uni}}
\and A. Trcka \inst{\ref{inst-UGent}}
\and E. M. Xilouris \inst{\ref{inst-Athens-Obs}}
\and N. Ysard \inst{\ref{inst-IAS}}
}

\institute{Sterrenkundig Observatorium, Universiteit Gent, Krijgslaan 281, B-9000 Gent, Belgium \label{inst-UGent} \\
\email{sebastien.viaene@ugent.be}
\and Centre for Astrophysics Research, University of Hertfordshire, College Lane, Hatfield AL10 9AB, UK \label{inst-UHerts}
\and National Observatory of Athens, Institute for Astronomy, Astrophysics, Space Applications and Remote Sensing, Ioannou Metaxa and Vasileos Pavlou GR-15236, Athens, Greece \label{inst-Athens-Obs}
\and Department of Astrophysics, Astronomy \& Mechanics, Faculty of Physics, University of Athens, Panepistimiopolis, GR15784 Zografos, Athens, Greece \label{inst-Athens-Uni}
\and Instituto de Radioastronom\'\i a y Astrof\'\i sica, UNAM, Campus Morelia, A.P. 3-72, C.P. 58089, Mexico \label{inst-UNAM} 
\and INAF -- Osservatorio Astrofisico di Arcetri, Largo E. Fermi 5, 50125, Firenze, Italy \label{inst-Firenze}
\and INAF -- Istituto di Radioastronomia, Via P. Gobetti 101, 40129, Bologna, Italy \label{inst-Bologna}
\and INAF -- Istituto di Astrofisica Spaziale e Fisica cosmica,  Via A. Corti 12, 20133, Milano, Italy \label{inst-Milano}
\and Space Telescope Science Institute, 3700 San Martin Drive, Baltimore, Maryland, 21218, USA \label{inst-STSCI}
\and School of Physics and Astronomy, Cardiff University, The Parade, Cardiff CF24 3AA, UK \label{inst-Cardiff}
\and Department of Physics and Astronomy, University College London, Gower Street, London WC1E 6BT, UK \label{inst-UCL}
\and Laboratoire AIM, CEA/DSM - CNRS - Universit\'e Paris Diderot, IRFU/Service d'Astrophysique, CEA Saclay, 91191, Gif-sur- Yvette, France \label{inst-Saclay}
\and Institut d'Astrophysique Spatiale, CNRS, Univ. Paris-Sud, Universit\'e Paris-Saclay,  B\^{a}t. 121, 91405, Orsay Cedex, France \label{inst-IAS}
\and Central Astronomical Observatory of RAS, Pulkovskoye Chaussee 65/1, 196140, St. Petersburg, Russia \label{inst-StPetersB-obs}
\and St. Petersburg State University, Universitetskij Pr. 28, 198504, St. Petersburg, Stary Peterhof, Russia\label{inst-StPetersB-uni}
}

\abstract{The star formation rate and the mass of interstellar medium (ISM) have a high predictive power for the future evolution of a galaxy. Deriving such properties is, however, not straightforward. Dust emission, an important diagnostic of star formation and ISM mass throughout the Universe, can be powered by sources unrelated to ongoing star formation. In the framework of the DustPedia project we have set out to disentangle the radiation of the ongoing star formation from that of the older stellar populations. This is done through detailed, 3D radiative transfer simulations of face-on spiral galaxies. We take special care in modelling the morphological features present for each source of radiation. In this particular study, we focus on NGC~1068, which in addition contains an active galactic nucleus (AGN). The effect of diffuse dust heating by AGN (beyond the torus) was so far only investigated for quasars. This additional dust heating source further contaminates the broadband fluxes on which classic galaxy modelling tools rely to derive physical properties. We aim to fit a realistic model to the observations of NGC~1068 and quantify the contribution of the several dust heating sources. Our model is able to reproduce the global spectral energy distribution of the galaxy. It matches the resolved optical and infrared images fairly well, but deviates in the UV and the submm. This is partly due to beam smearing effects, but also because the input dust distribution is not sufficiently peaked in the centre. We find a strong wavelength dependency of AGN contamination to the broadband fluxes. It peaks in the MIR, drops in the FIR, but rises again at submm wavelengths. We quantify the contribution of the dust heating sources in each 3D dust cell and find a median value of $83 \%$ for the star formation component. The AGN contribution is measurable at the percentage level in the disc, but quickly increases in the inner few 100 pc, peaking above $90 \%$. This is the first time the phenomenon of an AGN heating the diffuse dust beyond its torus is quantified in a nearby star-forming galaxy. NGC~1068 only contains a weak AGN, meaning this effect can be stronger in galaxies with a more luminous AGN. This could significantly impact the derived star formation rates and ISM masses for such systems.}


\titlerunning{Dust heating in NGC~1068}
\authorrunning{S. Viaene}

\maketitle

\section{Introduction}

Dust can be found in all star-forming galaxies, and even in a significant fraction of early-type galaxies \citep{DeLooze2010,Smith2012,diSeregoAlighieri2013}. Dust grains play a vital role in the formation of stars by regulating the gas temperature, catalysing the creation of molecules such as H$_2$ and CO, and subsequently shielding them. Due to this close connection with both stars and gas, dust is increasingly used as a tool to study the evolution of galaxies through time \citep[see][for a review]{Galliano2018}. Dust and gas are strongly intertwined in the interstellar medium (ISM) and usually follow a metallicity dependent gas-to-dust ratio \citep{Lisenfeld1998,Sandstrom2013,RemyRuyer2014,DeVis2019,Casasola2019}. As such, dust emission can not only be used to trace its own mass, but also the total mass of the interstellar medium \citep{Eales2012, Scoville2014, Hughes2017}. Accurate dust mass measurements are an important extra metric for evolutionary studies of ISM in galaxies, as neutral gas is hard to detect beyond the local Universe, and molecular gas tracers rely on empirical conversion factors \citep{Bolatto2013}. The most obvious effect of dust is of course its capacity to absorb and scatter UV and optical light. In the local Universe, roughly one third of starlight is absorbed by dust grains \citep{Skibba2011,Viaene2016, Bianchi2018}. This absorbed energy is re-emitted again at MIR-mm wavelengths where it can even be observed up to high redshifts with telescopes such as ALMA and \textit{Herschel}. Dust is therefore also a popular star formation tracer across cosmic time \citep[see e.g.][]{daCunha2010,Madau2014, Casey2014, Scoville2015}.

A big caveat of using dust as a tracer of either ISM mass or the star formation rate (SFR), is the assumed (often linear) conversion of FIR-submm emission into either of these quantities. At this point, it becomes important to understand how dust grains are heated to their equilibrium temperature. In practice, the most common assumption is that of a single heating source (e.g. star formation) producing a single radiation field, which is then parametrised with a single --average-- temperature component. Dust masses can then be estimated from a modified blackbody function \citep{Hildebrand1983, Bianchi2013} or a dust emission template \citep[e.g.][]{Dale2001, Draine2014,Jones2017}. In reality however, dust in galaxies is a mix of grain types and sizes, with multiple temperatures, and multiple heating sources. A more accurate determination of dust masses (and thus total ISM masses) requires more complex models for the dust emission. Such models also benefit the determination of the SFR from dust emission. In this case, it is important to only count the absorbed energy from new stars that is re-emitted by dust.

The main dust heating sources in galaxies have been qualitatively derived from correlations of dust emission with SFR or stellar mass tracers \citep{Galametz2010,Boquien2011,Smith2012b,Bendo2012,Foyle2013,Hughes2014} or from panchromatic models for a galaxy's spectral energy distribution (SED) \citep{Groves2012,Aniano2012,Dale2012,MentuchCooper2012,Ciesla2014,RemyRuyer2015,Boquien2016,Viaene2016}. 
The broadest of such studies was performed by \citet{Nersesian2019} as they modelled the global SED of 814 nearby galaxies of different morphologies. They found a wide variety in the individual contributions of dust heating sources, with a significant and sometimes dominant contribution of the old stellar populations. In a dedicated resolved study, \citet{Bendo2015} mapped the dust heating sources for 24 nearby face-on spiral galaxies. They found that in some galaxies the dust is predominantly heated by the on-going star formation, while some are heated by the old stellar population. Most galaxies, however, again show an mixed heating pattern with energy from both young and older stars contributing to the dust emission. 

In a pioneering study by \citet{DeLooze2014}, high-resolution 3D radiative transfer simulations were used to model the dust heating in M~51. The novelty in this study was to infer realistic, non-axisymmetric distributions for both stars and dust from observed broadband images. They found strong changes in the dust heating fractions for varying FIR/submm wavelengths, with an overall contribution of $37 \%$ from the old stellar populations. A similar model was created for the Andromeda galaxy (M~31) by \citet{Viaene2017a}. They found that the old stellar populations dominate the dust heating ($91 \%$ globally), even in the main star-forming ring of M~31. This effect was attributed to the large and old bulge of Andromeda. Most recently, M33 was also modelled in this way, finding old population heating contributions between $50\%$ and $80\%$ \citep{Williams2019}. In the context of the DustPedia project \citep{Davies2017}, we have set up a best-practice framework for such radiative transfer modelling and applied it to M~81 \citep{Verstocken2020} and to a set of nearby barred star-forming spirals (M~83, M~95, M~100 and NGC~1365; \citealt{Nersesian2020}).

All these studies assume a stellar origin for the energy absorbed by dust. In the radiative transfer models, three stellar populations are needed to match the observed panchromatic SED and images: an old (5-12 Gyr) stellar component, a non-ionising ($\sim 100$ Myr) component, and a population of ongoing obscured star formation. While this is acceptable for many star-forming spiral galaxies, there is evidence for other sources of dust heating. Dust grains can gain energy by capturing hot electrons, protons or even ions \citep{Jones2004}. In the latter two cases this can also partially destroy a grain (a process called sputtering). As such, cosmic rays produced by recent supernovae may be a secondary dust heating mechanism following recent star formation. In a similar way, hot gas can produce a flux of energetic electrons which directly heat the dust grains \citep{Bocchio2013}. 

Another potential dust heating source is the accretion disc around supermassive black holes. In galaxies hosting an AGN, the accretion disc certainly heats the dusty torus surrounding it and produces strong mid-infrared emission \citep[see e.g.][]{Fritz2006}. However, higher dust temperatures and anomalously bright nuclear FIR emission is found for galaxies hosting an AGN \citep{Wu2007,Bendo2012,Kirkpatrick2012,Verstappen2013, Kirkpatrick2015, Roebuck2016}. This circumstantial evidence suggests that the accretion disc may be powerful enough to heat ISM dust beyond the torus. \citet{Schneider2015} and \citet{Duras2017} investigated this phenomenon using radiative transfer simulations of quasars, and found that the AGN may contribute $30-70\%$ to the heating of diffuse dust. This phenomenon was confirmed through analysis of observations of a sample of QSOs by \citet{Symeonidis2016, Symeonidis2017}. While this may be an extreme case, it shows that AGN can contribute to the FIR-submm SED in galaxies. This effect influences the total mass and SFR estimates of galaxies hosting an AGN, but is not taken into account by SED fitting models.

The goal of this paper is to address the question of AGN-powered dust heating in the local Universe. We choose to take a quantitative approach to separate the contributions of star formation, old stellar populations and AGN. Therefore, we will focus on a single galaxy, but construct a detailed model based on 3D radiative transfer simulations. This is the fourth paper in a series of radiative transfer studies on individual galaxies. Within the DustPedia framework, the method was homogenized as much as possible to allow comparisons between the models. Paper I presents a radiative transfer model and dust heating analysis in the grand-design spiral M51 \citep{DeLooze2014}. Paper II \citep{Verstocken2020} is focussed on a streamlined modelling pipeline and its application to the spiral galaxy M81. A sample of barred galaxies was subsequently studied in paper III \citep{Nersesian2020}. There have been two spin-offs studies targeting the large Local Group galaxies M31 \citep{Viaene2017a}, and M33 \citep{Williams2019}. In the future we will also test this method on simulated galaxies to gain further insight in the dust heating mechanisms and identify opportunities to improve our method.

This paper focuses on NGC~1068 (M~77), a nearby ($D=10.1$ Mpc) Seyfert 2 SAb galaxy \citep{Osterbrock1993}. This AGN has no broad-line emission features, despite being a nearly face-on ($i = 28.1 \degree$) disc galaxy.  Instead, optical polarimetry reveals broad reflection lines \citep{Antonucci1985}. These observations suggest that the accretion disc is obscured by the torus for an observer on Earth \citep[see also][]{Marin2018}. VLBI measurements of a H$_2$O megamaser \citep{Greenhill1996} confirm that the torus is in fact seen edge-on. Recent ALMA observations of the core of NGC~1068 reveal a complex ISM, with the torus being inclined at only $34-66\degree$ at larger radii \citep{GarciaBurillo2016}. The orientation of the torus suggests that the accretion disc is in effect beaming its radiation straight into the galactic disc, which makes NGC~1068 a well-suited target to study AGN-powered dust heating. In addition, the galaxy is actively forming stars in its inner regions and hosts a relatively quiescent outer disc \citep{DAgostino2018}. We can thus also compare the relative importance of stellar heating mechanisms to each other, and to the AGN contribution.

In Sect.~\ref{sec:data} we describe DustPedia's panchromatic dataset for NGC~1068. The model preparation based on this data is outlined in Sect.~\ref{sec:modelling}. We describe our fitting process in Sect.~\ref{sec:fitting} and the best-fit high-resolution model in Sect.~\ref{sec:results}. The dust heating in NGC~1068 is analysed in Sect.~\ref{sec:dustHeating}. We finally discuss and summarise our findings in Sect.~\ref{sec:discussion}.

\section{Data} \label{sec:data}

This investigation relies mostly on broad-band imaging from the DustPedia\footnote{\url{http://dustpedia.astro.noa.gr}}  database \citep{Davies2017}. This comprehensive database contains the 875 nearby ($< 40$ Mpc) galaxies that were observed by the \textit{Herschel} Space Observatory. It includes uniformly reduced PACS \citep{PACS} and SPIRE \citep{SPIRE} maps, and broad-band images from major ground and space-based survey telescopes up to the UV. In addition aperture-matched photometry in the UV-mm is available. For more details on the image processing and photometry we refer to \citet{Clark2018}.

In particular, for NGC~1068, we make use of UV maps from GALEX \citep{GALEX}, optical images from SDSS \citep{SDSS}, MIR observations from WISE \citep{WISE}, and submm data from \textit{Herschel} \citep{Herschel}. We do not rely on the standard photometry of DustPedia in this case, as our model will focus on a smaller aperture. We thus recomputed the photometry measurements based on these new apertures (see Sect.~\ref{sec:modelling}). The AGN in NGC~1068 is especially bright in the MIR. Images from $\um{3.4}$ till $\um{24}$ are severely distorted by the PSF signature of their respective instruments. This virtually renders any morphological study impossible in this wavelength domain. We note that the images are not over-exposed in the MIR, meaning the integrated flux measurements are still useful and will serve as the most important constraint on the AGN model.

There are only two \textit{Spitzer}-IRAC bands ($\um{3.6}$ and $\um{4.5}$) for NGC~1068 in DustPedia. They unfortunately have strong blooming artefacts and do show signs of an over-exposed centre. We thus prefer the WISE images, which are of comparatively much better quality in this case, and cover a wider wavelength range ($\um{3.3}$ - $\um{22}$). The \textit{Spitzer}-MIPS bands do not have artefacts, but are still heavily PSF dominated (especially at $\um{24}$). Moreover, the PACS bands surpass them in spatial resolution. For these reasons, we do not include any \textit{Spitzer} observations in our dataset. 

The 2MASS observations (also in the DustPedia database) turned out to be too shallow to capture the entire disc of NGC~1068. We instead use the UKIDSS \citep{UKIDSS} $H$ and $K$ band observations which are deeper and have better spatial resolution. Most importantly, the $H$ band is the reddest band without a significant PSF signature and will be vital to determine the spatial distribution of the old stellar populations. For a similar reason, we will also use the narrow-band $H\alpha$ map of NGC~1068 \citep{Knapen2004}. The map will be an important constraint on the spatial distribution of the ongoing star formation in the absence of useable $\um{22}$ or $\um{24}$ images. We discuss this in more detail in the next section.

\begin{figure*}
	\includegraphics[width=\textwidth]{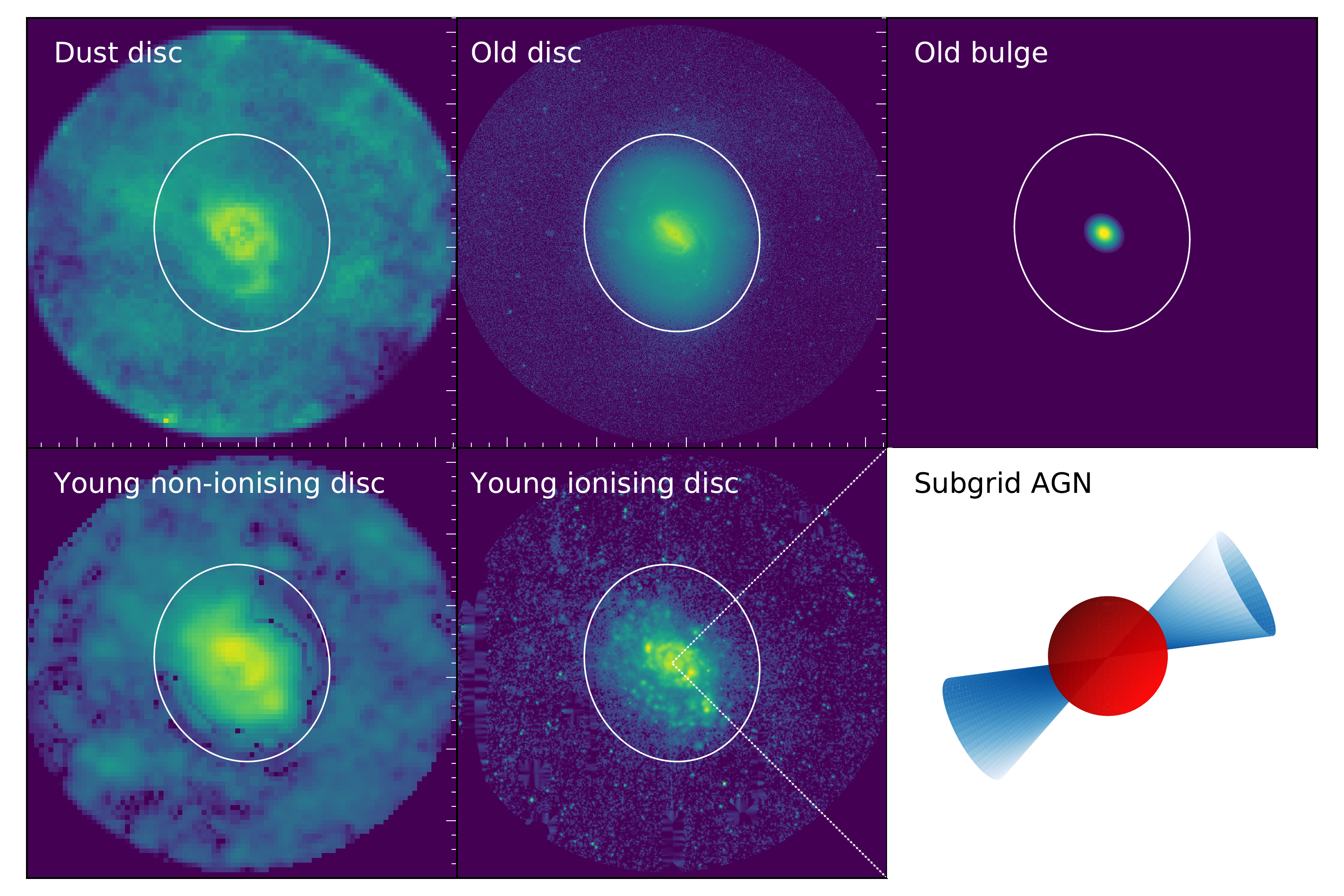}
    \caption{2D representations (on the plane of the sky) used as input morphologies for the 3D geometries for the model of NGC~1068. The old bulge and AGN model are based on analytical prescriptions while the other morphologies are derived from observed imagery. The model $\chi^2$ is only evaluated within the white ellipse to ensure a sufficient signal-to-noise ratio. The ellipse has a semi-major axis of 83 arcsec or 4 kpc. The AGN cone orientation in the bottom right panel is arbitrary and only for display purposes.}
    \label{fig:geometries}
\end{figure*}

\begin{figure}
	\includegraphics[width=0.45\textwidth]{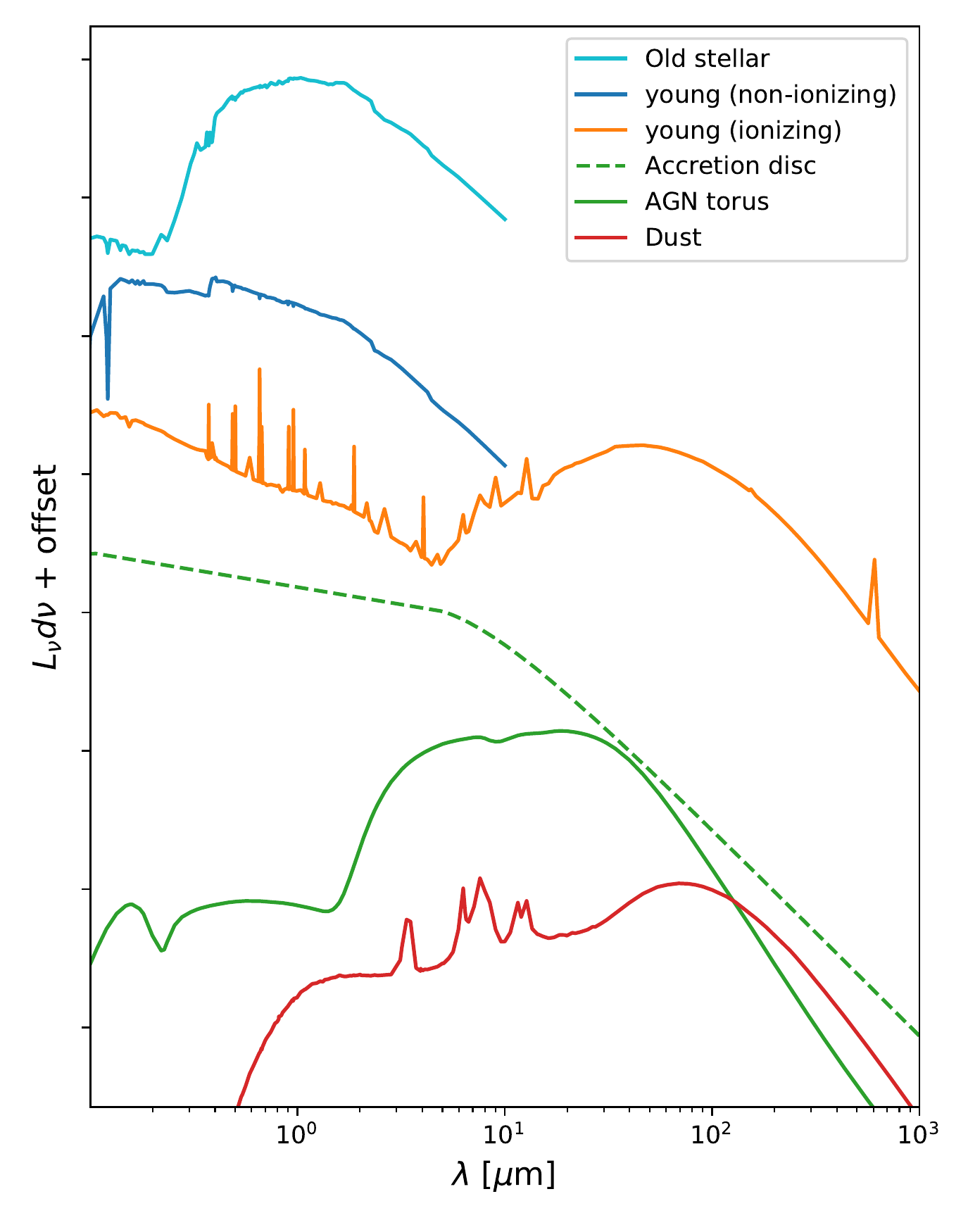}
    \caption{Overview of the different SED templates adopted in the model. Templates are scaled for better visualization.}
    \label{fig:sedcomponents}
\end{figure}

\section{Model preparation} \label{sec:modelling}

Our modelling strategy closely follows the prescriptions outlined in \citet{Verstocken2020}, which, in turn were based on the work by \citet{DeLooze2014}. The requirements for a panchromatic, 3D and resolved radiative transfer fit are three stellar components and one dust component. Here, we limit ourselves to a brief description at each modelling step and highlight where we deviate from \citet{Verstocken2020}. 

\subsection{Dust component}

The dust component requires a 3D geometry, plus the choice of a dust mix with associated optical properties. While similar studies in the literature mostly used analytical density profiles as input geometries, it is now possible to derive them from observations directly. This allows the use of non-axisymmetric and more clumpy distributions for stars and dust, leading to more realistic models.

The most obvious way to trace the dust morphology is through a map of the dust mass surface density. However, this requires resolved SED fitting up to the SPIRE $\um{350}$ band and drastically limits the working resolution. However, the input dust distribution does not need to be normalised as the total dust mass will be a free parameter. We therefore construct a map of $FUV$ attenuation, which also traces the dust distribution but importantly attains higher spatial resolution \citep{DeLooze2014}. The per-pixel determination of $A_{FUV}$ is the same for all galaxies studied in this series. It relies on the calibrations of \citet{Cortese2008}. They derive the attenuation based on the total-infrared to $FUV$ ratio ($TIR/FUV$) with second-order adjustments based on the observed $FUV-r$ colour. The $A_{FUV}$  map is shown in Fig.~\ref{fig:geometries} (top left panel) labelled as the dust disc. To convert this to a 3D density distribution, we deproject the 2D image along the minor axis and stretch the emission in each deprojected pixel out into the vertical ($z$-axis) direction according to an exponential profile. We set the scale height of this exponential profile to $76$ pc, which is half that of the old stellar scale height as suggested by \citet{DeGeyter2014}. For the spectral properties of dust absorption, scattering and emission we adopt the THEMIS dust model for diffuse Milky Way dust \citep{Ysard2015,Jones2017}, in line with other DustPedia studies. A typical emission SED of this dust model is shown in Fig.~\ref{fig:sedcomponents}.

\subsection{Stellar components}

Each of the stellar components has a 3D spatial distribution associated with them, and a template shape for their SED. The templates are simple stellar populations taken from the \citet{Bruzual2003} library assuming a \citet{Chabrier2003} initial mass function (see the blue lines in Fig.~\ref{fig:sedcomponents}). We split the stellar components into three age groups. The old stellar population contains stars that are much older than $100$ Myr, and therefore no longer contribute to the classic SFR tracers \citep{Kennicutt2012}. Typically, one can trace this component best with either WISE $\um{3.4}$ or IRAC $\um{3.6}$ observations. However, in the case of NGC~1068, these images are dominated by the PSF shape and do not provide clear morphological information. We therefore rely on the UKIDSS $H$ band to trace the old stellar component. The additional advantage of using this shorter waveband is that the AGN contamination in the centre is minimal. We perform a bulge/disc decomposition on the $H$ band image to further split the geometry into a Sérsic bulge and a disc. The bulge model (the only analytical density profile) is already 3-dimensional by construction. The disc is deprojected in the same way as the dust map, but now given an exponential scale height of $151$ pc, which is $1/8.26$ of the scale length as suggested by \citet{DeGeyter2014}. The second stellar component is a young non-ionising stellar population with average age of $100$ Myr and is based on the attenuation-corrected GALEX-$FUV$ map. This map was deprojected and given an exponential scale height equal to the dust scale height.

\begin{figure*}[!h]
	\includegraphics[width=\textwidth]{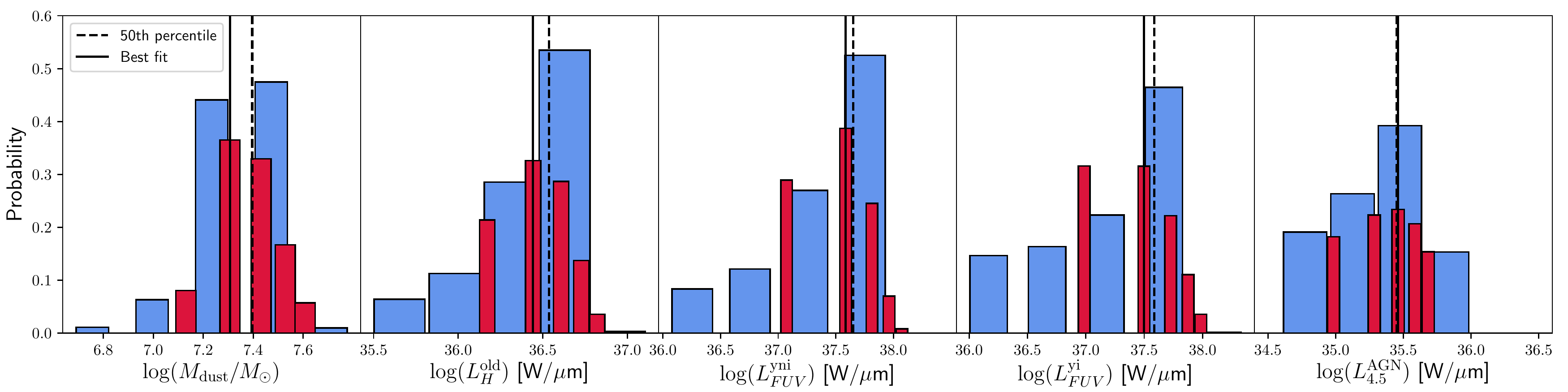}
    \caption{Probability density distributions for the free parameters in our model configuration. The spacing between bars represent the spacing of the parameter grid. The width of the bars is set so they do not overlap due to the unequal spacing. The blue distribution corresponds to the first batch of models. The parameter grid was refined based on this batch. After the second fit iteration, the red distributions are found. The median values of the second batch are indicated with vertical dashed lines. The solid vertical lines indicate the values for the model with the lowest overall $\chi^2$.}
    \label{fig:probabilities}
\end{figure*}

The third stellar component models the ongoing star formation in the form of young ($10$ Myr) ionising stars still in their birth clouds. The MAPPINGS III \citep{Groves2008} model of stars and dust in a birth cloud with this average age was used as SED template for this component (see Fig.~\ref{fig:sedcomponents}). To trace the ongoing star formation, we use the narrow-band continuum-subtracted H$\alpha$ emission. This image still has an unusually bright nucleus. \citet{DAgostino2018} showed that the AGN is responsible for $25 \%$ of the H$\alpha$ emission in the galaxy, and for $42 \%$ of the [NII] line emission. Both emission lines fall within the narrowband filter image used here. To mitigate the contamination, we mask out the nuclear H$\alpha$ peak in the inner $350$ pc or $7$ arcsec \citep[see also][for a similar approach]{Casasola2015}. The \texttt{astropy} routine \texttt{interpolate\_replace\_nans}\footnote{\url{http://docs.astropy.org/en/stable/convolution/}} was used to infer the H$\alpha$ emission from the surrounding disc. Note that this procedure preserves the exponential flux profile of the H$\alpha$ disc. The total flux in the image is reduced by $41\%$ after this correction, which is at the high end of the range estimated by \citet{DAgostino2018}. The resulting image is converted to a 3D geometry by deprojecting it and assuming an exponential scale height of $38$ pc, which is half that of the dust scale height and represents the embedded nature of the star-forming regions. 

\subsection{AGN component}

An important novelty is the addition of an AGN in our model. Radiative transfer simulations of accretion discs surrounded by a dusty torus have been done in the past \citep[see e.g.][]{Nenkova2008,Stalevski2012}. However, given the enormous difference in scale between the torus and the galactic disc, it is not necessary to solve the transfer of radiation in the same simulation. As such, we construct a subgrid model for the AGN in NGC~1068. The two AGN emission components (the accretion disc and the dusty torus) are both modelled as point-like sources located at the centre of the galaxy. 

The accretion disc spectrum is modelled as a composition of power laws with indexes varying as a function of wavelength. We used the \citet{Schartmann2005} prescription, which combines observed and theoretical evidence and is commonly used as a heating source in radiative transfer models for dusty tori in AGN (Fig.~\ref{fig:sedcomponents}). Its emission peaks at UV-optical wavelengths and its normalisation is tied to the total torus luminosity. A part of the radiation of the accretion disc is blocked locally by dust within the torus. We have implemented this as an anisotropic (cone-shaped) emission profile with an opening angle of 20$\degree$ around the polar direction (see Fig.~\ref{fig:geometries}, bottom right panel). This simple geometrical implementation mimics the same internal absorption fraction as our preferred choice of torus model (see below): only $6 \%$ of the accretion disk light escapes the torus. The orientation of the accretion disc is such that it beams directly into the galactic disc, along the minor axis (inclination $i=0\degree$, azimuth $\phi=90\degree$), with the obscured part pointing to the observer, as expected for a Type-2 Seyfert galaxy with a hidden Type-1 core like NGC 1068 \citep{Marin2018}.

We assume the torus emission to be isotropic, which is a fair approximation given our limited spatial resolution (44 pc) and the fact that its MIR radiation is far more optically thin than the obscured accretion disc (see also Fig.~\ref{fig:geometries}, bottom right panel). To choose an emission model from the dusty torus, we have performed a fit to the photometric points used by \cite{LopezRodriguez2018}, using the the upgraded model grid of \cite{Fritz2006}, presented in \cite{Feltre2012}. The former is a collection of spatially-resolved data that are dominated by the torus emission and includes crucial features such as the near-infrared emission and the $\um{9.7}$ silicate feature, observed in slight absorption for this AGN (Fig.~\ref{fig:sedcomponents}). The model that provides a good fit to these point has a dust full-opening angle of $140^\circ$, an outer-to-inner radius of 30. The equatorial optical depth is 6, measured at $\lambda=\um{9.7}$, and the dust density $\rho(r,\theta)$ varies within the torus according to:
\begin{equation}
\rho(r,\theta)=\alpha r^\beta  e^{- \gamma |\cos (\theta )|}
\end{equation}
where $\alpha$ is directly related to the value of the equatorial dust optical depth, and the two parameters $\beta$ and $\gamma$ regulate the radial and height density profile (see Eq.~3 in \cite{Fritz2006} for further details). They assume the value of 0 and 4, respectively, for the best fit model.

The AGN model is normalised by the $\um{4.6}$ luminosity since this wavelength is particularly sensitive to the presence of very hot dust. It is not the goal of this paper to further constrain the torus or accretion disc properties. We simply require a realistic subgrid implementation for the AGN. As such the total luminosity is the only free parameter of the AGN component (with a fixed ratio between accretion disc and torus luminosity).

\section{Radiative transfer SED fitting} \label{sec:fitting}

The different components described in the previous section define the framework for our model of NGC~1068. Radiative transfer simulations are computationally demanding, so the free parameters in this model need to be limited. We therefore rely on well-informed estimates for the spatial and spectral distribution of each component as outlined above. For the dust density distribution the only free parameter is the total dust mass, $M_\text{dust}$. For the other components, the free parameters are the normalisation luminosities in certain wavebands: The spatial and spectral distribution of the young non-ionising (yni) and young ionising (yi) stellar populations are normalised by their $FUV$ luminosity: $L^\mathrm{yni}_{FUV}$ and $L^\mathrm{yi}_{FUV}$, respectively. The old stellar components are normalised together (assuming a fixed bulge-to-disc ratio) by their total luminosity in the $H$ band:  $L^\mathrm{old}_{H}$. Finally, the subgrid AGN SED is normalised by the total emission of accretion disc and torus at $\um{4.6}$: $L^\mathrm{AGN}_{4.6}$ . Here again the ratio between disc and torus luminosity is fixed to guarantee the internal energy balance. In total, the model thus has five free parameters. 

We make use of SKIRT \citep{Baes2011, Camps2015} to perform the radiative transfer calculations in the 3D simulation space. The stellar and AGN components serve as sources of radiation. The dust component acts as a sink of UV-optical-NIR radiation, which is then reprocessed and emitted at longer wavelengths. Light is also anisotropically scattered by the dust grains.  SKIRT can perform these simulations in a highly efficient way thanks to multiple optimisation techniques for the transfer of radiation towards the observer \citep{Baes2011}, effective quantisation of the dust density distribution using a binary tree dust grid \citep{Saftly2014}, and support for various types of parallel computing \citep{Verstocken2017}. 

Our goal is to find the optimal combination of free parameters that reproduce the observations. Despite the efficiency of SKIRT, this is not a trivial task. We therefore run a suite of low resolution simulations and use the global broadband SED of NGC~1068 as observational constraint since it is efficient to generate the synthetic, line-of-sight model fluxes. The radiation transfer is evaluated in a set of 3D dust cells of variable volume, but roughly equal in mass. With a dust grid of $1.3 \times 10^6$ cells we reach an effective resolution of $44$ pc in each dimension. In each simulation, the model is evaluated in $134$ wavelength bins by $1 \times 10^6$ photon packages per wavelength. Doing so, each set of parameters can be evaluated in under one hour (on average) on a 16 CPU node.

With each point in the 5D parameter space taking 16 CPU hours, it becomes impractical to run a classic nonlinear fitting algorithm. We therefore adopt the same two-step fitting approach as for the other DustPedia-modelled galaxies \citep{Verstocken2020, Nersesian2020}. In a first run, we explore a wide parameter space, sparsely sampled around a well-motivated initial guess (see below). For each parameter set, the observed and model broadband SED is compared by summing the $\chi^2$ metric. The fluxes are measured in 2D (sky projection) images inside an elliptical aperture with semi-major axis of 4 kpc (see Fig.~\ref{fig:geometries}). Each wavelength regime (UV, optical, NIR, MIR, FIR and submm) is given equal weight to the total squared sum \citep[see][for a detailed description]{Verstocken2020}. The parameter values for each fit can then be weighted by $\exp(-\chi^2/2)$ to sample the probability density distribution. 

As an initial guess, we use the global properties derived by \citet{Nersesian2019} which are based on SED modelling with CIGALE \citep{CIGALE}. They find
\begin{equation}
\begin{cases}
M_\mathrm{dust}^\mathrm{CIGALE} = (1.71 \pm 0.23) \times 10^7 \, M_\odot \\ \\
\mathrm{SFR}^\mathrm{CIGALE} = 13.4 \pm 3.6 \, M_\odot\mathrm{yr}^{-1} \\ \\
A^\mathrm{CIGALE}_{FUV} = 3.64 \pm 0.31
\end{cases}
\end{equation}
This dust mass can directly be used as an initial guess. From the SFR, we compute an initial guess for the UV luminosity of the ionising stellar populations. The $A_{FUV}$ can be used to de-redden the observed $FUV$ luminosity, which we assume here for simplicity as the sum of the ionising and non-ionising stellar components. Furthermore, we use the observed $H$ band luminosity as an initial guess for the total (bulge+disc) old stellar populations as this band is relatively free from other contributions. For the AGN torus, our initial guess normalisation was set at $2.50\times 10^{35}$ W $\um{}^{-1}$ and tied to this an accretion disc luminosity $0.15 \times 10^{35}$ W $\um{}^{-1}$. These luminosities provide the best fit of our model to the estimated fluxes of the central 20 pc from \citet{LopezRodriguez2018}. The initial parameter set for the model is listed in Table~\ref{tab:parameters}.

For galaxies without a strong AGN, radiative transfer fits with three parameters ( $M_\mathrm{dust}$, $L_{FUV}^\mathrm{yni}$ and $L_{FUV}^\mathrm{yi}$) are sufficient. The challenge of adding an AGN is that the MIR SED becomes even more convoluted and difficult to decode into separate contributions. To deal with this added complexity we first introduce $L_{4.6}^\mathrm{AGN}$ as a free parameter to scale the AGN template SED. Secondly, we also allow the old stellar population template SED to scale up and down. This is set by $L_{H}^\mathrm{old}$ since the $H$ band is a good tracer of this stellar population and the contamination of the AGN is expected to be minor. An adequate coverage of this parameter space thus requires many more sample points. We build a sparse but wide 5D Cartesian parameter grid around our initial guesses. The grid spans more than one order of magnitude in $M_\mathrm{dust}$, $L_{4.6}^\mathrm{AGN}$ and $L_{H}^\mathrm{old}$, and more than two orders of magnitude in $L_{FUV}^\mathrm{yni}$ and $L_{FUV}^\mathrm{yi}$. Each dimension is uniformly sampled in log space with 5 points during this first run, which already amounts to $3125$ simulations.

The resulting probability distributions of the first batch are shown in Fig.~\ref{fig:probabilities} by the blue histograms. These have a clear peak for all free parameters which importantly does not occur at the edge of the sampled space. This means that we are already able to put coarse constraints on all parameters. The median values (50th percentile of the PDFs) are summarised in Table~\ref{tab:parameters}.

\begin{figure*}[!h]
	\includegraphics[width=\textwidth]{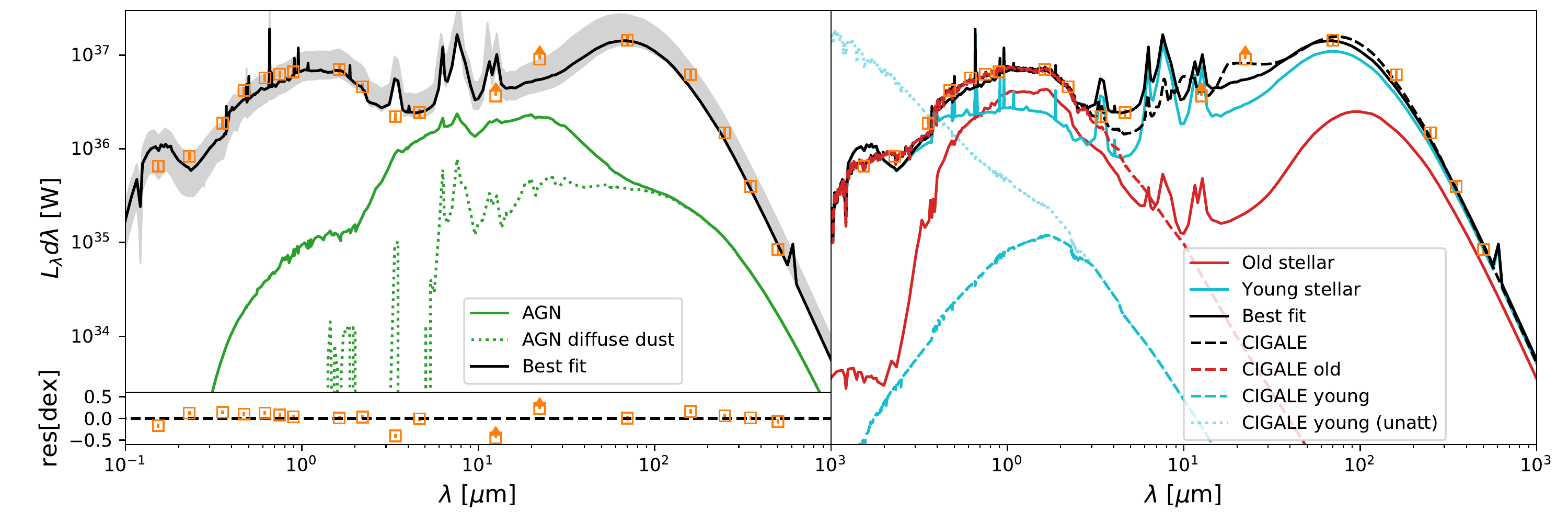}
    \caption{SEDs from the best-fit model for NGC~1068. Left: the global SED (black line) fits the observed (orange) broadband fluxes. The gray shading indicates the 16th to 84th percentile area around the fit. Triangles indicate lower limits. The corresponding residuals are shown in the bottom panel. The solid green line shows the total emission for a model with only the AGN + dust component. The dotted green line then corresponds to the light reprocessed by the diffuse dust only in this simulation. Right: same best-fit SED and data points. The solid blue and red lines correspond to simulations where only one component (young or old, respectively) is present alongside the dust. A CIGALE model is also shown (dashed black line). This model can be decomposed into young and old populations (blue and red dashed lines, respectively). The intrinsic young population (blue dotted line) of the CIGALE fit is also shown.}
    \label{fig:SED_full}
\end{figure*}

\begin{table*} 
\caption{Overview of the free parameters of our model setup together with their initial guess values. After each fitting batch, the 50th percentile corresponds to the median value for that fit. The uncertainties on the 50th percentiles reflect the width of the corresponding bin. The last column lists the parameter values for the model with the lowest overall $\chi^2$.}
\label{tab:parameters}
\centering     
\begin{tabular}{lcccc}
\hline \hline
  Parameter & Initial guess & Batch 1 (50th pct) & Batch 2 (50th pct) & Best fit \\
\hline
$\log(M_\mathrm{dust}/M_\odot)$           & 7.23   & $7.43_{-0.19}^{+0.15}$   & $7.40_{-0.12}^{+0.11}$   &   7.31   \\
$\log(L_{H}^\mathrm{old}/W \um{}^{-1})$  	& 36.65 & $36.64_{-0.25}^{+0.17}$ & $36.54_{-0.19}^{+0.15}$ &   36.44 \\
$\log(L_{FUV}^\mathrm{yni}/W \um{}^{-1})$	& 37.56 & $37.85_{-0.32}^{+0.22}$ & $37.65_{-0.28}^{+0.21}$ &   37.58 \\
$\log(L_{FUV}^\mathrm{yi}/W \um{}^{-1})$ 	& 37.46 & $37.76_{-0.33}^{+0.22}$ & $37.59_{-0.31}^{+0.23}$ &   37.50 \\
$\log(L_{4.6}^\mathrm{AGN}/W \um{}^{-1})$	& 35.47 & $35.63_{-0.32}^{+0.30}$ & $35.41_{-0.24}^{+0.16}$ &   35.41 \\
\hline
\end{tabular}
\end{table*}

Based on the distributions of the first batch, we can narrow the parameter space significantly and adopt a finer sampling. We ensure that over $90\%$ of the probability is captured in the parameter space for the second batch. We again use 5 sample points per parameter, but now uniformly distributed in linear space. The probability distributions for this second batch of 3125 simulations are also shown in Fig.~\ref{fig:probabilities} (red histograms). Each distribution shows a clear peak, with the exception of $L_{FUV}^\mathrm{yi}$. It seems we are reaching the limit of precision by which we can constrain this parameter. The PDF for $L_{4.6}^\mathrm{AGN}$ is also rather broad for the second batch.  The parameter values for the best-fitting model (lowest $\chi^2$) lie close to the median parameters (50th percentile values) and are listed together in Table~\ref{tab:parameters}. These values also correspond well with our initial guess. We explore the 2D probability distributions in Appendix~\ref{app:cornerplot} and find no strong degeneracies or suspicious features in these figures. We conclude that at this point the free parameters of our model are reasonably constrained. Adding another iteration would again require significant computational resources and not significantly reduce the uncertainties. In the next section, we analyse the best-fitting model in further detail.

\section{A 3D high-resolution model for NGC~1068} \label{sec:results}

\begin{figure*}
	\includegraphics[width=0.97\textwidth]{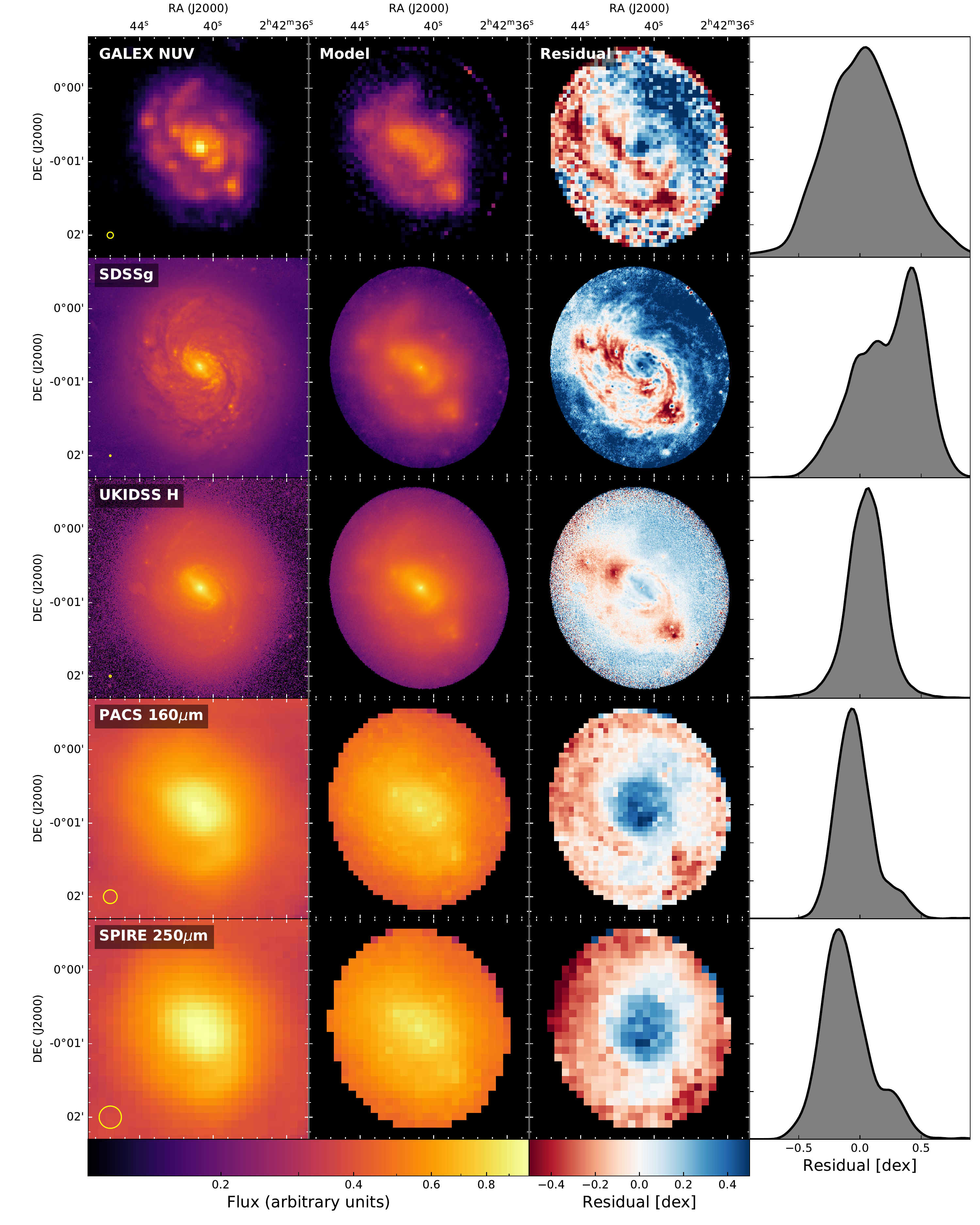}
    \caption{Spatial comparison of our best-fit model to observed broad-band images (left column). The second column contains the model images. Note that the model images are not convolved by any PSF and only hold the intrinsic model PSF (a composite of the input images). Residual images are shown in the third column. Corresponding residual distributions are represented by KDE plots in the right column. The area under the KDE curves is normalized to 1.}
    \label{fig:residuals}
\end{figure*}

\subsection{SED comparison}

The best-fit model obtained in the previous section forms the basis for our dust heating analysis. We first produce a higher quality version of the best-fit model by shooting five times as many photons per wavelength ($5\times10^6$), and almost twice the number of wavelength points ($252$). This increases the spectral sampling and reduces the inherent Monte Carlo noise in the dust cells and in the resulting images. To separate the contributions of each source of radiation, each simulation is run again with the same dust distribution, but only one source component each time: the old (bulge and disc), young (non-ionising and ionising discs) and AGN (accretion disc and torus) component, respectively. This process takes another 1280 CPU hours. The resulting SED, along with its uncertainty levels, is shown in Fig.~\ref{fig:SED_full} (left panel, black line).

Globally, our model follows the observed SED fairly well. The main stellar and dust peak match the datapoints closely and within the uncertainties. There are, however, points of discrepancy. First, the $FUV$ point is overestimated by the model, while the $NUV$ point is underestimated. This issue has been noted before in galaxies with a significant contribution of the young ionising templates \citep{DeLooze2014, Verstocken2020,Nersesian2020}. This signature in the template for ionising stellar populations is amplified by a broad bump in the THEMIS attenuation curve. \citet{Nersesian2020} compared the attenuation curve (normalised to $A_V$) of our model to their galaxies. They found that NGC~1068 exhibits the flattest attenuation curve of the sample. It also lies closest to the extinction curve corresponding to the THEMIS diffuse Milky Way dust model, which is also used here in our radiative transfer simulations.	

Also driven by the dust model are the strong aromatic peaks in the MIR SED in  Fig.~\ref{fig:SED_full}. The adopted emission model was calibrated on diffuse dust in the Milky Way. The strength of MIR emission features is often reduced in AGN compared to star-forming galaxies \citep[see e.g. ][]{Sales2010} although this may be attributed to dilution by the bright MIR continuum \citep{AlonsoHerrero2014}. MIR spectra of NGC~1068 do reveal line emission even close to the nucleus \citep{Mason2006, Howell2007} due to ongoing star formation there. Environmental changes in dust composition and grain mixture  are capable of significantly changing the dust emission profile \citep{Koehler2015} but this is currently not possible in our model. Destruction of the smallest amorphous hydrocarbon (a-C) grains would for example reduce the $\um{3}-\um{13}$ emission band intensities. At the same time, a lack of a-C grains would also reduce the UV bump and provide a better match to the GALEX data points. This underlines the need to allow for dust evolution in the next generation of radiative transfer codes. A framework for such simulations is currently being developed (Camps in prep.). We did not wish to skew our model by fitting to wavebands that contain MIR emission peaks and therefore do not include the WISE $\um{3.4}$ and $\um{12}$ flux in the $\chi^2$ weighting.

As already mentioned in Sect.~\ref{sec:data}, good quality maps of NGC 1068's main disc are lacking in the MIR. At $\um{12}$ and $\um{22}$ especially, the PSF signature (a blend of the disc and AGN emission) is larger that the field of view we consider. A such we treat these two points as lower limits. The most discrepant point is the WISE $\um{22}$ flux, which is underestimated by the model. We looked into models with a stronger AGN contribution, but these lift at the same time also the MIR continuum. Changing the ratio between the stellar components also did not improve the fit without introducing discrepancies at other wavelengths. The observed WISE $\um{22}$ flux in our aperture is higher than the MIPS $\um{24}$ flux (not shown and also suffering from strong PSF signatures). Part of this discrepancy could thus be attributed to the observational uncertainty, but still our model seems unable to capture the exact SED shape in the MIR. This may partly be due to the subgrid implementation of the torus. A non-isotropic torus model or a more clumpy one could boost the emission at $\um{22}$ in the AGN SED \citep[see][]{Feltre2012}. However, as indicated earlier, it is not the goal of this paper to further constrain the torus properties themselves, but instead we focus on how radiation escaping the AGN affects the host galaxy.

Alternatively, changing one of the stellar emission templates could bridge the discrepancy.  The MIR emission in particular can be boosted by tweaking the young ionising component template. The covering factor used in the \citet{Groves2008} template  library can be increased to boost the emission from warm dust surrounding star-forming regions, while at the same time keeping the escaping UV radiation relatively low. Unfortunately, introducing yet another free parameter in the optimisation procedure would be computationally difficult. In addition, one of the main goals of this paper series is to provide a suite of radiative transfer models for different galaxies with consistent set of fixed parameters \citep{Verstocken2020}. Only the template luminosities and dust mass are considered free parameters, which allows us to investigate heating mechanisms across the sample \citep{Nersesian2020}.

We also run separate simulations containing only one emission component: the young (non-ionising+ionising) stellar populations, the old (bulge+disc) stellar populations and the AGN. Each of these simulations still contain the dust component and are simply re-runs of the best-fit simulation with fewer emission sources. The diffuse dust is now heated solely by the components that are present in the simulation. We note that the sum of these SEDs will not add up to the global SED (with all components) as dust emission is a non-linear effect.

In the left panel of Fig.~\ref{fig:SED_full} we plot the SED of the simulation with the AGN (accretion disc+torus) + dust only. It only contributes significantly to the MIR, but does not dominate the global SED in any regime. An important observation here, however, is the broad tail of the AGN SED in the submm. This is not direct emission from the AGN, but light reprocessed by the diffuse dust in the galaxy (see green dotted line in Fig.~\ref{fig:SED_full}). This already indicates that the AGN radiation is influencing the surrounding ISM. However, the effect is rather small as the dust emission lies typically one order of magnitude below the emission generated by the old or young components. We compute a bolometric luminosity $L_\mathrm{AGN}^\mathrm{bol} = 0.4 \times 10^{37} W$, compared to $L_\mathrm{dust}^\mathrm{bol} = 2.7 \times 10^{37} W$. For the total observed SED, the bolometric luminosity is $L_\mathrm{tot}^\mathrm{bol} = 4.3 \times 10^{37} W$. Following \citet{Bianchi2018} we can compute $f_\mathrm{abs} = L_\mathrm{dust}^\mathrm{bol}/ (L_\mathrm{tot}^\mathrm{bol} - L_\mathrm{AGN}^\mathrm{bol})= 0.69$ the fraction of light that is reprocessed into dust emission. Our estimate corresponds well with the $0.68$ they find from a CIGALE model of the global SED.

The SEDs of the old+dust and young+dust simulations are plotted in the right-hand panel of Fig.~\ref{fig:SED_full}. The dominance of the young+dust simulation (blue line) is immediately clear. Its SED lies well above the SED of the other simulations in the UV, MIR and FIR regime. This is already an indication that the FIR flux is powered by ongoing star formation. The second most luminous SED comes from the old+dust simulation, although it only really dominates in the NIR. Still it seems to contribute some energy to the submm emission.

We can compare our model and subcomponents with 1D SED models like the ones fitted by CIGALE. This method is well suited for fast modelling of statistical samples, but as a trade-off lacks 3D information and radiative transfer effects. We adopt the modeling method of \citet{Nersesian2019} and apply this to the fluxes of NGC 1068 in the ellipse region under consideration (see Fig.~\ref{fig:geometries}). The resulting SEDs of the total and subcomponents of the CIGALE model are shown as dashed lines in Fig.~\ref{fig:SED_full}. The global SED fits the observed data very well. The amount of free parameters (and their priors) in CIGALE is sufficient to alleviate the discrepancies present in our radiative transfer model. Especially interesting is the emergence of a second peak in the MIR to match the $\um{22}$ point. This does seem to go at the cost of a lower continuum at shorter wavelengths, underestimating the $\um{4.6}$ point.

A breakup of the CIGALE model into subcomponents provides more insight in the global fit. Fig.~\ref{fig:SED_full} shows that the observed SED model is dominated by the old component. This can be compared to the old stellar component of the radiative transfer model, bearing in mind that the latter SED also contains dust emission. The CIGALE subcomponents shown here do not include dust emission. Similarly, we can compare the young components and find a vast discrepancy between the models. The CIGALE young component has a negligible contribution to the observed SED. To inspect this mode closely we also plot the unattenuated SED of the young component (blue dotted line). The unattenuated SED is much more luminous pointing towards extreme levels of attenuation. This is likely related to the emission peak at $\um{22}$. As CIGALE tries to fit this data point, it needs to boost the continuum and PAH luminosity, which requires strong attenuation of UV and optical light. It is important to note that this model does not contain an AGN. Additional freedom of the AGN component could help fitting the MIR without skewing the attenuation for the young population.

From this comparison we conclude that 1D SED fitting tools can provide good fits to the global fluxes of galaxies hosting an AGN. However, caution is needed in interpreting these models. In large samples it is a priori not clear which galaxies require a low-luminosity AGN component (like NGC 1068) in the model. Looking at the subcomponents provides useful insight in this case.Standard CIGALE fits with extremely attenuated young stellar component and a double infrared bump can identify AGN in a sample of galaxies. This initial result motivates more comparisons between 3D RT SED fits and 1D tools to highlight deficiencies in both methods and act upon them.

\subsection{Image comparison}

To further assess the quality of our model, we generate synthetic images. We convolve the 3D datacube produced by SKIRT using the spectral response curves of five representative broadband filters across the SED. We did not convolve the images with a corresponding PSF as it is difficult to determine the model kernel. The PSFs of the images that were used to extract the input density distributions, as well as the resolution of the dust grid, result in a complex effective PSF of the model image. As such we simply compare the spectrally convolved model images with their observed counterparts in Fig.~\ref{fig:residuals} and indicate the instrument PSF as a yellow circle. We also produce residual maps in dex units as $\log_{10}(\text{observation}) - \log_{10}(\text{model})$, and plot the kernel density estimates (KDE) for the pixel distributions in these residual maps.

The model images reproduce the observed counterparts within 0.5 dex. Visually the correspondence is quite good in the UV, optical and NIR maps. However, the residuals reveal significant asymmetries corresponding to the spiral arms. This is in part due to the deprojection of 2D input images to create a 3D density distribution, where light is smeared out in the vertical direction. On top of that, we do not resolve the dust extinction at sufficient resolution, which introduces residual features at smaller scales. The effect is worst in the SDSS $g$ band image, where the spatial resolution is the highest. In this band the flux is also systematically underestimated by the model as highlighted by the KDE plot. In the NIR, there are two strong positive residual features on opposite sides of the galaxy. They correspond to star-forming regions which are over-luminous in the model (represented by the ionising component). Globally, though, the NIR regime shows the best correspondence with the observations.

The model FIR images do not show the bright central area, but the PACS map does peak sharply in the nucleus. This can mainly be attributed to the difference in PSF, but the model also underestimates the flux in the centre (while matching the total flux in the FIR regime). We speculate that this may be due to an insufficient central dust density in the input map. The current dust geometry map is based on pixel-by-pixel determinations of $A_{FUV}$. This relies on the calibrations of \citet{Cortese2008}, which derive the stellar age bin from the $FUV-r$ colour. It is possible that the AGN emission in NGC~1068 can locally skew these age estimates, resulting in a less reliable $A_{FUV}$ estimate in the centre of the galaxy. We have tried to fit a model where the dust geometry followed more closely the PACS $\um{70}$ map. This dust map is more centrally peaked than the current one based on $A_{FUV}$ and presented in Appendix~\ref{app:dustmaps}. However, this alternative map causes severe attenuation in the UV and optical bands, too much dust emission in the FIR and a less favourable fit to the observed global fluxes. We therefore did not analyse this model further and stick to the model based on the $A_{FUV}$ dust map. Knowing the qualities and caveats to the best-fitting model, we analyse the dust heating sources and energy balance in NGC~1068 in the next section.

\subsection{Light fractions}

\begin{figure}
	\includegraphics[width=0.50\textwidth]{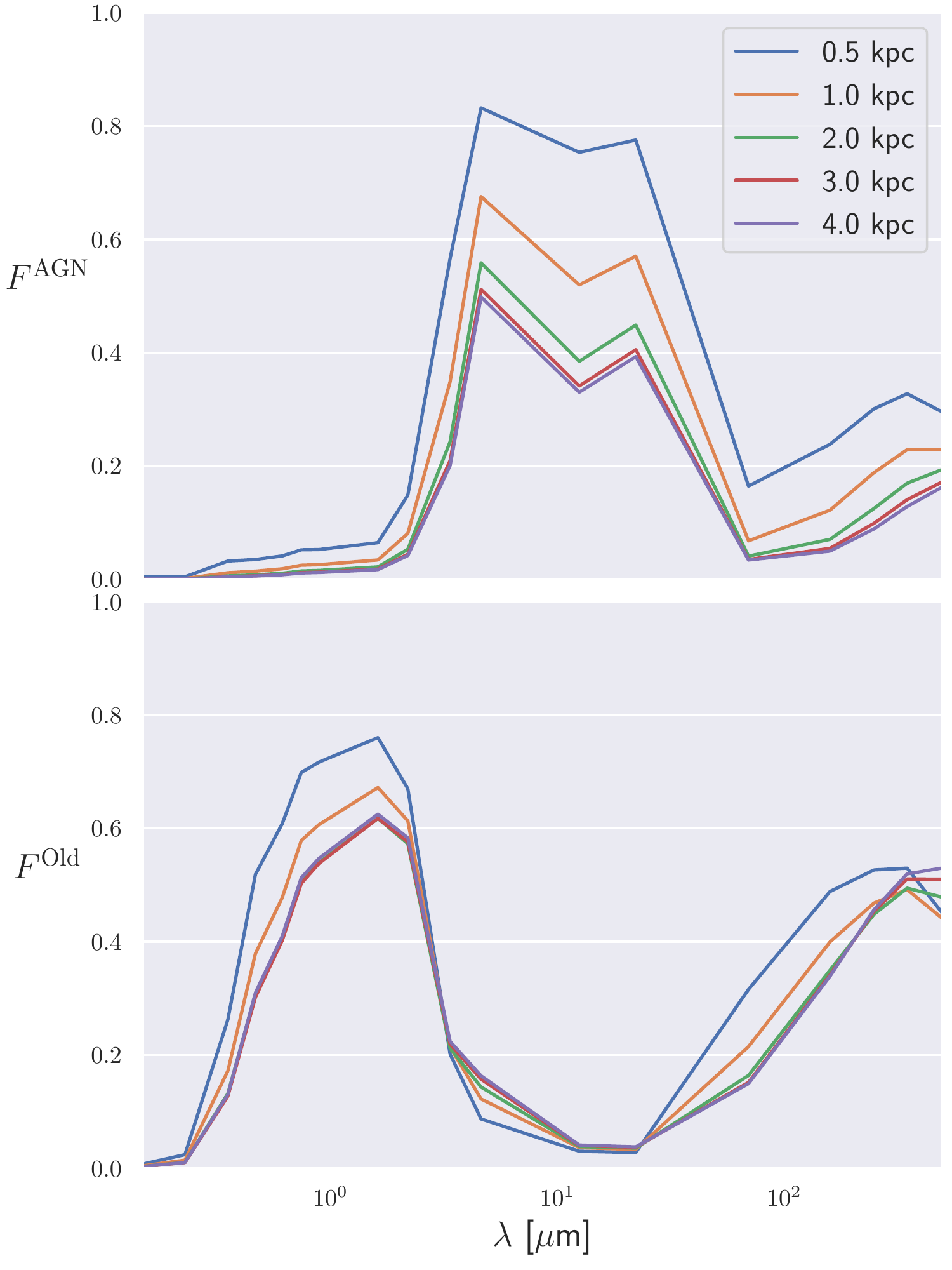}
    \caption{Fraction of emission originating from the AGN component to the total emission per band, $F^\text{AGN}$ (top panel) and from the old component $F^\text{Old}$ (bottom panel). Different lines correspond to different concentric elliptical apertures. The light fractions are derived from the integrated light within the full aperture.}
    \label{fig:agnlight}
\end{figure}

The global SED of our model further allows us to estimate the light contamination of sources unrelated to star formation in NGC~1068. We therefore compute, in each band, the emission ratio between the AGN-only and the total simulation (still including the same dust component), $F^\text{AGN}_\lambda$ and similarly for the old stellar population: $F^\text{Old}_\lambda$. It is useful to quantify these contributions since several broadband colours or luminosities are used to directly derive physical properties of galaxies. An adequate correction for star-forming galaxies containing an AGN is necessary before applying such recipes. Likewise, a correction for light coming from older stellar populations can improve estimates of star-formation and attenuation curves \citep[see e.g.][]{Roebuck2019}.

Fig.~\ref{fig:agnlight} (top panel) shows $F^\text{AGN}_\lambda$ for all bands in our dataset, and for several concentric elliptical apertures. As expected, there is a general increase in the AGN light ratio towards the innermost regions of the galaxy. $F^\text{AGN}_\lambda$ averages to only $0.12$ inside a 4 kpc major axis radius and across the wavelength range. In the innermost aperture (0.5 kpc), this average increases to $0.32$. Along the wavelength axis, all apertures exhibit the same behaviour. There is virtually no AGN contamination in the UV bands and only a minor contamination in the optical bands ($F^\text{AGN}_\lambda = 0.01-0.05$). There is a strong increase in the NIR and the peak $F^\text{AGN}_\lambda$ is reached in the WISE $\um{4.6}$ band and varies from $0.50$ within 4 kpc to $0.83$ within the inner 0.5 kpc. The peak actually extends well into the MIR, but drops off quickly towards the PACS $\um{70}$ band. In the FIR, an interesting positive slope points to an increasing contribution with increasing wavelength. The intrinsic AGN (disc+torus) emission model does not have this behaviour, as is evident from the right panel in Fig.~\ref{fig:SED_full}. This is actually an indirect effect as AGN light is reprocessed by the diffuse dust. It is important to realise that even SPIRE bands can thus be contaminated by the AGN.

The radial and spectral variation in $F^\text{Old}_\lambda$ is displayed in the bottom panel of Fig.~\ref{fig:agnlight}. There is less variation between the different apertures than for the AGN light fraction. Still a similar pattern emerges: the highest fractions are found in the center and can be attributed to the bulge population of older stars. Along the spectral axis, interestingly, there are two peaks. The first peak around $\um{1}$ is expected and corresponds to the peak of direct emission of the old component. The second peak lies around $\um{350}$ where $F^\text{Old}_\lambda$ still reaches $0.5$. This can be attributed to reprocessed light from the old stellar populations by the diffuse dust. We note that this number does not suggest a dust heating fraction of $50 \%$ because the conversion of dust emission SED to the input energy is non-linear. We study the input energies further in the next section but the light fractions already qualitatively confirm that both old stellar populations and the AGN affect the total dust emission in this galaxy.

\section{Dust heating sources} \label{sec:dustHeating}

The main advantage of 3D radiative transfer simulations is the inherent information on the energy balance in every dust cell. Our model, based on NGC~1068, contains the absorbed energy in each cell and thus summarises how the dust in that cell is heated before re-emitting this energy at longer wavelengths. In line with our previous work \citep{Verstocken2020,Nersesian2020}, we make use of the total (wavelength-integrated) absorbed energy per stellar component and per volume element (dust cell). We derive three quantities: $f_\text{young}$, the combined heating fraction of the young (non-ionising + ionising) component,  $f_\text{old}$ the heating fraction of the old stellar component, and $f_\text{AGN}$, the AGN heating fraction. They reflect dust heating through ongoing star formation, the general radiation field and through AGN emission, respectively, and are defined as 
\begin{equation}
\begin{array}{l}
f_\text{young} = \displaystyle \frac{L^{abs}_\text{yni} + L^{abs}_\text{yi}}{L^{abs}_\text{tot}}, \\
f_\text{old} = \displaystyle \frac{L^{abs}_\text{old}}{L^{abs}_\text{tot}}, \\
f_\text{AGN} = \displaystyle \frac{L^{abs}_\text{AGN}}{L^{abs}_\text{tot}},
\end{array}
\end{equation}
where the $L^\text{abs}_\text{k}$ refers to the absorbed luminosity for component k. These quantities are computed per dust cell. The total absorbed luminosity is the sum of all components per dust cell: $L^{abs}_\text{tot} = L^{abs}_\text{yni} + L^{abs}_\text{yi} +L^{abs}_\text{AGN} +L^{abs}_\text{old}$.

\begin{figure*}
	\includegraphics[width=0.93\textwidth]{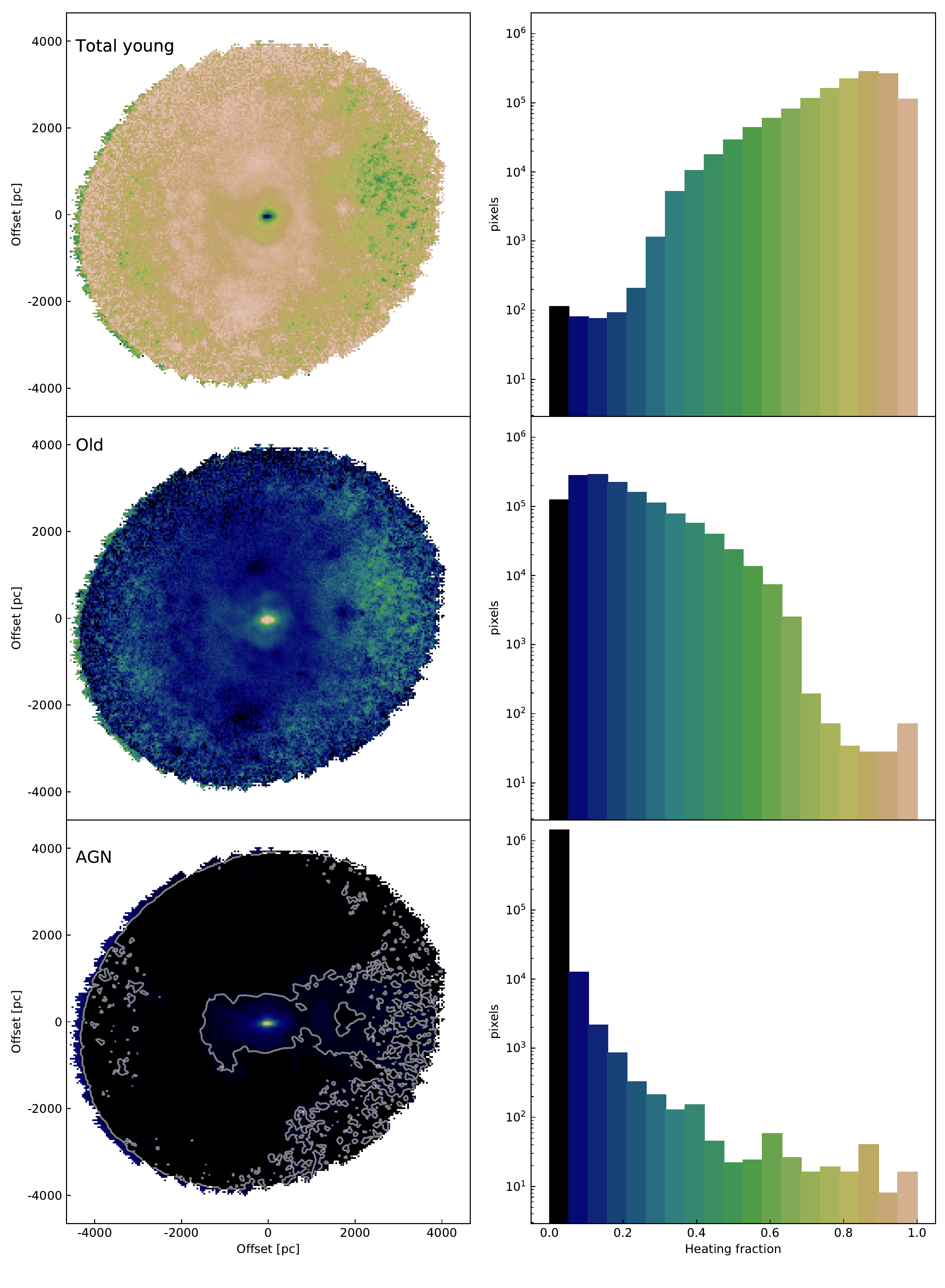}
    \caption{Dust heating fractions in a slice through the mid-plane of the face-on view of our model (left panels). The direction of the AGN beam is aligned with the horizontal axis of the figure. The colour code is set on the histograms on the right. The histograms contain all dust cells (including the ones above and below the mid-plane). Top row corresponds to $f_\text{young}$, middle row to $f_\text{old}$ and bottom row to $f_\text{AGN}$. The white contour corresponds to the $0.5\%$ level in $f_\text{AGN}$ and highlights an asymmetry in the heating of dust in the disc.}
    \label{fig:midplaneheating}
\end{figure*}

Fig.~\ref{fig:midplaneheating} summarises the dust heating fraction of each component. The left panels show a cut through the mid-plane of dust cells in a face-on view of the galaxy model. In this view, the direction of the AGN beam is aligned with the horizontal axis of the figure. The colour coding is set by the histograms on the right. The histograms contain all dust cells (including the ones above and below the mid-plane). Globally, a median $f_\text{young} = 0.83$ was found, which underlines the dominance of star formation as a heating source in this galaxy model. In contrast, the old stellar populations have a median  $f_\text{young} = 0.16$ and the AGN effect is minor with a median $f_\text{AGN} = 0.006$.

More detail is visible on a local level in Fig.~\ref{fig:midplaneheating}. The young component peaks in a ring around the centre, and in two zones above and below it. These structures can also be distinguished in optical images of NGC~1068 and are associated with enhanced star formation (see the ionising disc map in the bottom middle panel of Fig.~\ref{fig:geometries} or the $NUV$ map in Fig.~\ref{fig:residuals}). The base level of $f_\text{young}$ remains high across the disc, but drops below $0.5$ in some patches at larger radii. This is where the ongoing star formation fades and the radiation field created by the old stellar population takes over in heating the dust. In many ways, the map of $f_\text{old}$ appears as the complement of the $f_\text{young}$ map. This indicates that stars (young or old) are still very much the dominant heating sources in this galaxy.

Quite remarkable is the central dip in $f_\text{young}$, which can not be fully explained by the peak in $f_\text{old}$ due to the bulge. It spatially correlates with the peak in $f_\text{AGN}$ in the bottom left panel of Fig.~\ref{fig:midplaneheating}. The central $\sim 0.5$ kpc is the main zone of influence of the AGN. Here, the AGN-powered dust heating even dominates that of the surrounding star-forming disc. The $f_\text{AGN}$ map also shows a clear directional preference for the AGN heating (roughly along the horizontal axis of Fig.~\ref{fig:midplaneheating}). The main direction is guided by our implementation of the AGN accretion disc, beaming straight into the disc with a conical opening angle of $20\degree$ (see Sect.~\ref{sec:modelling}). The map also shows a slight left-right asymmetry (highlighted by the white contour), where the AGN heating can be noted further out towards the right side of the map. The star-forming disc fades off more quickly in that direction, leaving room for the AGN to increase its contribution to the dust heating.

\begin{figure}
	\includegraphics[width=0.50\textwidth]{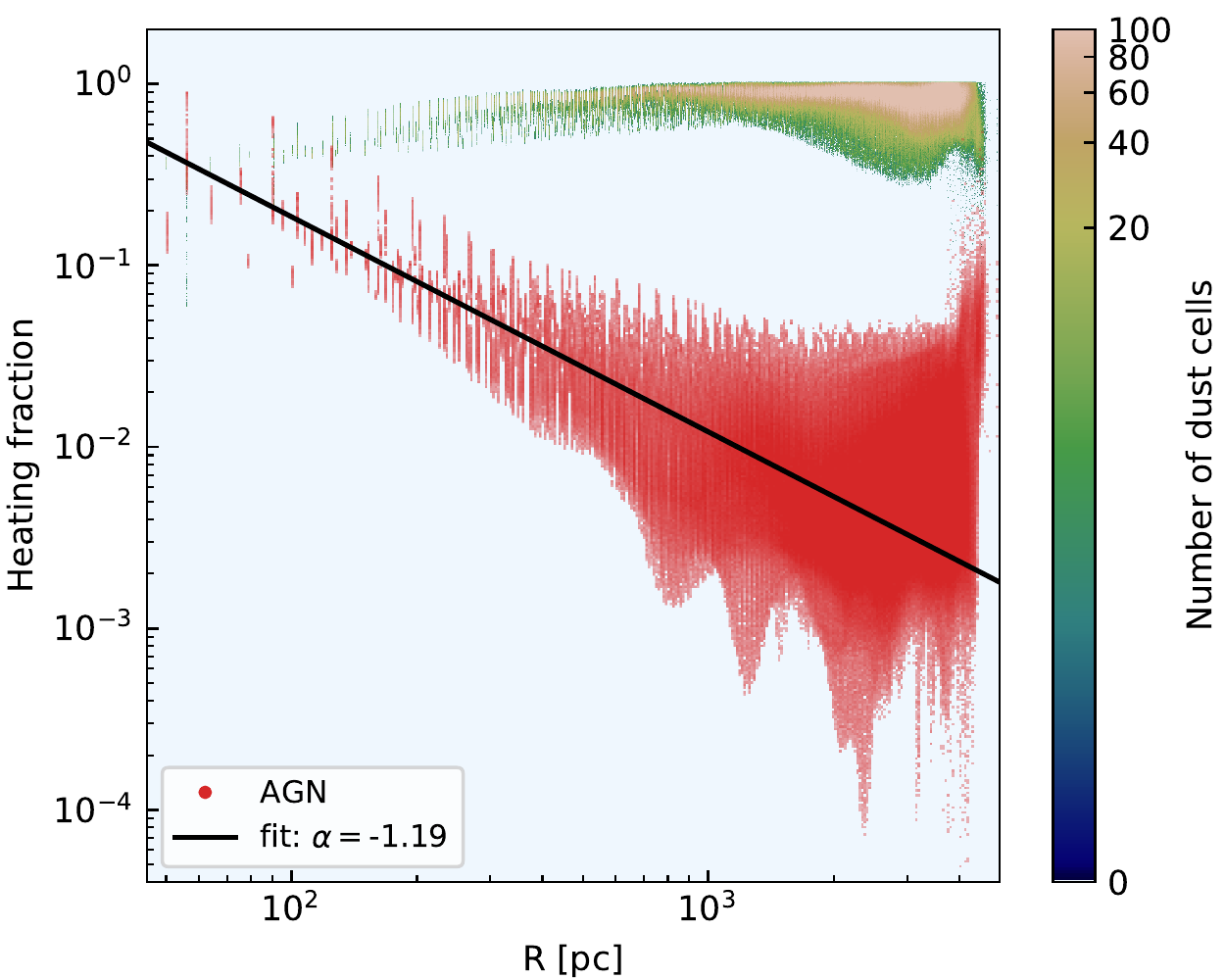}
    \caption{Scatter density plots showing the radial profiles of the dust heating fraction. The colour bar corresponds to the density in points of $f_\text{young}$. The red points correspond to $f_\text{AGN}$. The power-law fit of Eq.~\ref{eq:powerlaw} is shown as a black line.}
    \label{fig:radialprofiles}
\end{figure}

To further quantify the AGN-powered dust heating, we plot the radial profiles of $f_\text{AGN}$ and $f_\text{young}$ in Fig.~\ref{fig:radialprofiles}. The same trend as in Fig.~\ref{fig:midplaneheating} arises. The ongoing star formation dominates the dust heating in most of the dust cells in the disc with fractions between $60$ and $95 \%$ and with appreciable scatter at every radius. The radial distribution for $f_\text{AGN}$ exhibits a steep decline from the centre towards the outskirts. We can approximate the radial decline of the AGN heating fraction quite well with a power law:
\begin{equation} \label{eq:powerlaw}
f_\text{AGN} =  0.0117 \left( \frac{R}{\text{kpc}} \right)^{-1.19}.
\end{equation}
In the above fit we excluded data points inside the inner $50$ pc to avoid limited sampling and our spatial resolution limit. We also excluded the points beyond $3000$ pc to avoid edge-effects. The power law thus only holds within these radial limits. More models of galaxies hosting an AGN are required to verify whether this power-law behaviour is universal or not. Our analysis suggests that AGN-powered dust heating is appreciable ($> 5 \%$) inside the inner few 100 pc, and remains at the percentage-level throughout the galaxy.

\section{Discussion and summary} \label{sec:discussion}

We have created a radiative transfer model based on observed imagery of NGC~1068. This galaxy contains an extended old disc and bulge, and a smaller star-forming disc with spiral arms. NGC~1068 also bears an AGN and its accretion disc appears to be edge-on while the galaxy is roughly face-on. As such the accretion disc beams directly into the star-forming disc. This particular configuration makes the galaxy an ideal target to investigate a scenario where the AGN heats diffuse dust beyond its torus. While there is circumstantial evidence for such AGN-powered dust heating in the Local Universe  \citep{Wu2007,Bendo2012,Kirkpatrick2012,Verstappen2013,Kirkpatrick2015, Roebuck2016}, the phenomenon was thus far only simulated for quasars \citep{Schneider2015,Duras2017}. For such objects, observational evidence suggests that the AGN heating fraction is at least $5-10 \%$ \citep{Symeonidis2016, Symeonidis2017}. QSOs are, however, extreme environments with AGN that are orders of magnitude more luminous that in the local universe. To our knowledge, we present the first full 3D radiative transfer fit to a local star-forming galaxy with an AGN.

The complex geometry of structure of NGC~1068 makes it difficult to disentangle the main components of our model. We used linear combinations of observed broad and narrow-band images to obtain the purest possible 2D maps of old, non-ionising, and ionising stellar populations. The main challenge was to construct a dust mass map at a workable resolution. We deviated from the standard recipes provided by \citet{Verstocken2020} as we could not use bands that are dominated by the instrumental PSF signature (notably the MIR). Our model can reproduce the global SED of NGC~1068 and many of the resolved features (see Fig.~\ref{fig:SED_full} and Fig.~\ref{fig:residuals}). However, it is less successful in retrieving the FIR wavebands due to a slight shortage of dust mass towards the centre. Adding more dust in these areas would increase the heating fraction of the AGN and would produce a less favourable fit to the global SED. With these caveats in mind, our model should be seen as a representation of a star-forming galaxy, based on NGC~1068. The subsequent dust heating results are therefore still realistic. \\

Our best-fit radiative transfer model has a dust mass and and SFR (derived from young stellar luminosities), which are in which are in line with a classic CIGALE SED fit (Table~\ref{tab:parameters} and \citealt{Nersesian2019}). However, major differences are found between the relative importance of the attenuated stellar components. In the CIGALE model, the young stellar population is extremely attenuated in order to boost the MIR emission and compensate for the lack of AGN in the model. When modeling larger samples of galaxy SEDs it is thus worthwhile to check the subcomponents of 1D SED models. Signatures similar to the one found here could point to a low-luminosity AGN which requires more tailored SED models. This is an example of how 3D RT models can inform 1D SED models of large samples even though it is not efficient to perform a RT fit for each individual galaxy.

The AGN in our model has a bolometric luminosity of  $L_\text{bol}^\text{AGN} = 3.97 \times 10^{43} \, \text{erg}\,\text{s}^{-1}$, which is somewhat lower than previous AGN models for NGC~1068 \citep[e.g. ][]{LopezRodriguez2018}. This could in part be due to our assumption of an isotropic point source for the subgrid implementation of the AGN torus. The AGN in NGC~1068 is in any case quite weak for Seyfert 2 galaxies \citep[see e.g.][]{Lusso2012}. It is thus possible that the reported AGN dust heating fractions are on the low end of the potential spectrum. Still, even for this relatively weak AGN, we find percentage-level dust heating out to 4 kpc from the centre. The AGN heating fraction declines radially following a powerlaw with index $-1.19$. In the inner 500 pc, the heating fraction rises quickly above $10 \%$ and peaks above $90 \%$ in the inner resolution element of our simulation ($40$ pc). This is also reflected in the contribution of AGN light to each broadband flux in NGC~1068 (Fig.~\ref{fig:agnlight}). Even in the SPIRE bands there is contamination in the emission due to reprocessed AGN energy.

Globally, we find that most of the dust in the NGC~1068 model is heated by ongoing star formation. The median of this fraction is $83 \%$ when all dust cells are considered. This corresponds to a reduction of the $L_\text{dust}$-derived SFR from $11.7 \, M_\odot\text{yr}^{-1}$ in the full model to $8.4 \, M_\odot\text{yr}^{-1}$ in the model where only young (non-ionising and ionising) stellar populations heat the dust. These estimates were obtained using the \citet{Kennicutt2012} conversion from  $L_\text{dust}$ to SFR, which adopts the IMF from \citet{Kroupa2003}. Relatively speaking, this is a fairly small correction as NGC~1068 is the galaxy with the highest $f_\text{young}$ in our project sample (containing M~81 and four barred spirals M~83, M~95, M~100 and NGC~1365). This is clearly shown already in Fig.~8 of \citet{Nersesian2020}, where we plotted $f_\text{young}$ as a function of sSFR (the SFR divided by stellar mass). The dust cells of our model lie in line with the general increasing trend that other star-forming galaxies exhibit. Here, SFR and stellar mass were computed per dust cell, already taking into account the fact that not all dust emission arises from the young stellar populations.

Most of the additional energy to heat the dust actually comes from the old stellar populations $\sim 16 \%$ in our model, with only a negligible part from the AGN (on global scales). This is also in line with the global energy balance study of the DustPedia sample \citep{Bianchi2018}, where AGN do not stand out from the rest of this sample of local galaxies. However, Fig.~\ref{fig:midplaneheating} clearly shows that there are significant local differences. In the extreme case, considering the dust cells in the inner 100 pc, the median heating fractions average out at $34 \%$ and $28 \%$ for the young and for the AGN heating sources, respectively. The remaining energy can be attributed to the small bulge of old stars. FIR-based SFR indicators or dust mass estimates which include this part of the galaxy are thus susceptible to significant bias both due to the AGN and to the radiation field generated by the old stellar populations.

This study is the third in a series after \citet{Verstocken2020} and \citet{Nersesian2020} where we investigated the effect of dust heating in face-on large galaxies in the DustPedia database. The novelty of this study is that we added an AGN in the model as a source of radiation. We have pushed our models to the limit by requiring high-resolution images from FUV to submm wavelengths, and by fitting 3D radiative transfer simulations to these data. We are thus for the first time able to actually quantify the dust heating fraction for separate heating sources in a sample of galaxies. Our main conclusion is that star formation alone is inadequate to explain the observed FIR emission. In all cases, a significant contribution of the old stellar population is needed. 

In the particular case of NGC~1068, a low-luminosity AGN is also required to match the observations. In systems hosting a stronger AGN or in quasars, the AGN-powered heating of the diffuse dust will be even higher. Hence, simply removing a (fitted) AGN torus component from the FIR flux will still lead to an overestimation of the SFR of the host galaxy.

These findings, in particular the necessity for multiple heating sources, should be considered when building the next generation of galaxy SED modelling tools.

\begin{acknowledgements}
DustPedia is a project funded by the EU (grant agreement number 606847) under the heading "Exploitation of space science and exploration data (FP7-SPACE-2013-1)" and is a collaboration of six European institutes: Cardiff University (UK), National Observatory of Athens (Greece), Instituto Nazionale di Astrofisica (Italy), Universiteit Gent (Belgium), CEA (France), and Université Paris Sud (France). \\ 
We like to thank the anonymous referee for the constructive feedback and suggestions that helped improve this paper. \\
SV gratefully acknowledges the support of the UGent-BOF fund. SV and MB acknowledge the support of the Flemish Fund for Scientific Research (FWO-Vlaanderen). \\
The computational resources (High-Performance Cluster) and services used in this work were provided by the University of Hertfordshire (UK). \\
This research made use of Astropy,\footnote{http://www.astropy.org} a community-developed core Python package for Astronomy \citep{astropy:2013, astropy:2018}. 

\end{acknowledgements}

\bibliographystyle{aa} 
\bibliography{allreferences}

\appendix

\section{2D probability distributions} \label{app:cornerplot}

\begin{figure*}
	\includegraphics[width=\textwidth]{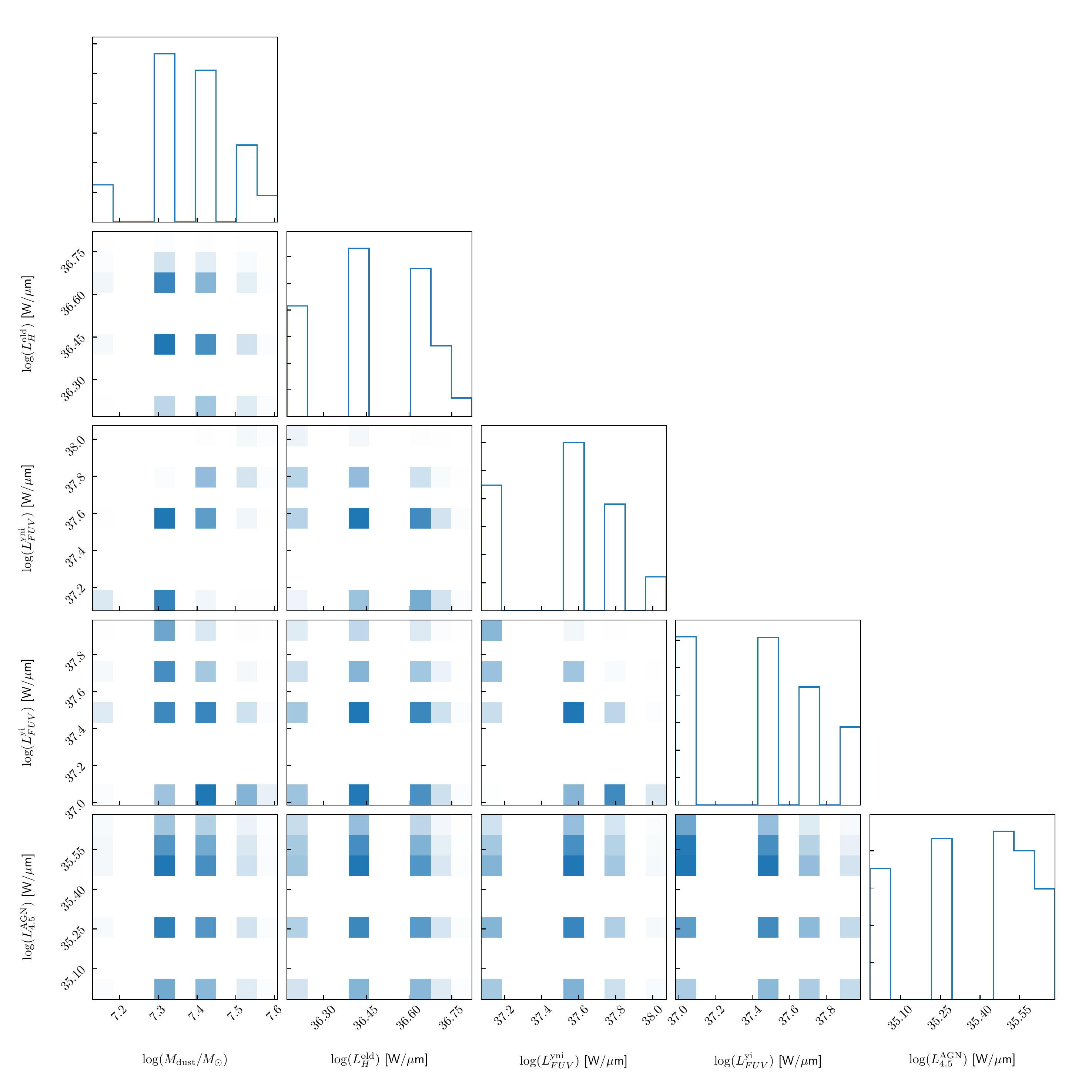}
    \caption{Probability distributions of our model parameters ($M_\mathrm{dust}$, $L_{H}^\mathrm{old}$, $L_{FUV}^\mathrm{yni}$, $L_{FUV}^\mathrm{yi}$ and $L_{4.6}^\mathrm{AGN}$) for the second optimization iteration (red histograms in Fig.~\ref{fig:probabilities}). Darker shades of blue indicate higher probability. The histograms show probability associated with the parameter on the abscissa, with the total probability in the bins normalized to 1.}
    \label{fig:cornerplot}
\end{figure*}

In Fig.~\ref{fig:probabilities} the probability distributions of the individual parameters are shown. To explore the probability space further, we can look at 2D probability distributions. Fig.~\ref{fig:cornerplot} shows this in the form of a corner plot\footnote{Produced using the python \texttt{corner} package \citep{corner}}, where the 1D histograms correspond again to the ones in Fig.~\ref{fig:probabilities}. Because our parameter grid only samples five values per parameters, the 2D probability distributions are only sparsely sampled, making it ineffective to compute probability contours. It is therefore difficult to draw strong conclusions from this plot, however some features can be noted.

The luminosity of the old stellar component ($L_{H}^\mathrm{old}$) is  well constrained with no significant link to the other parameters. We also don't see any correlation between the AGN luminosity and the other properties, although it is evident that that the probabilities for $L_{4.6}^\mathrm{AGN}$ are spread out more evenly across the parameter space. This is different for $M_\mathrm{dust}$, $L_{FUV}^\mathrm{yni}$ and $L_{FUV}^\mathrm{yi}$, which do show vague correlations among each other. This points to a mild degeneracy between the UV luminosities of the young stars and the dust component. This is not surprising because they influence the observed UV flux in opposite ways. In principle, this degeneracy could be broken by perfect sampling of the infrared SED. Since we only sparsely sample this regime, some residual correlation is possible between these properties. However we can still sufficiently constrain each parameter individually.

\section{Alternative dust map} \label{app:dustmaps}

\begin{figure}[!ht]
	\includegraphics[width=0.45\textwidth]{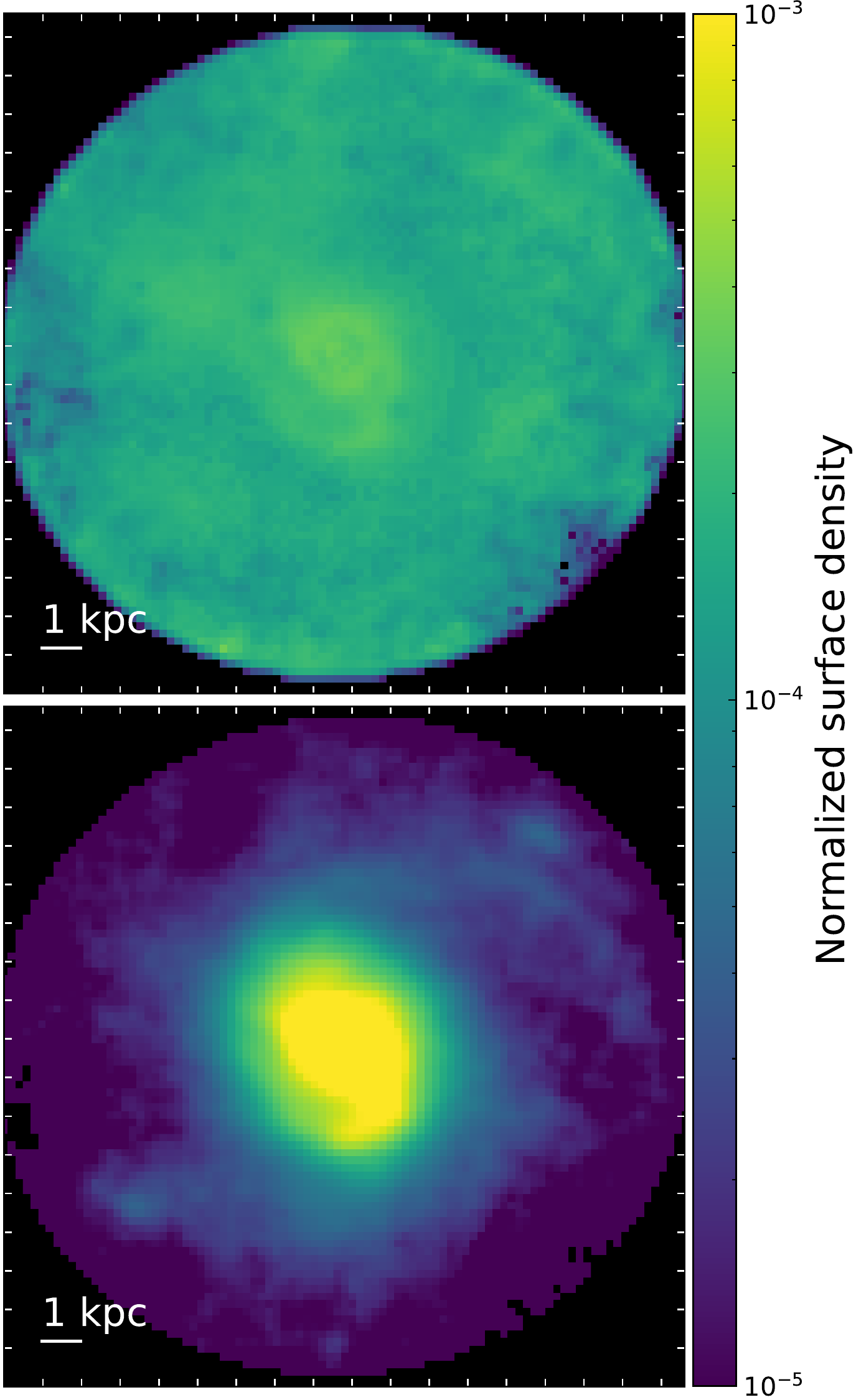}
    \caption{2D representations (on the plane of the sky) of the dust mass surface density used as input morphologies for the 3D dust distribution. Top panel is the same A$_{FUV}$ based map as in Fig.~\ref{fig:geometries} but on a logarithmic colour scale. Bottom panel is the alternative dust mass map based on the PACS $70 \, \mu$m image, which peaks strongly in the center.}
    \label{fig:dustmaps}
\end{figure}

Our standard model for NGC~1068 includes a dust component that is based on a 2D dust mass surface density map. This map was derived from the per-pixel $A_{FUV}$ estimate following the recipe of \citep{Cortese2008}. This recipe is based on $TIR/FUV$ and $FUV/r$ colours. It is possible that the AGN affects these measurements, resulting in a less reliable $A_{FUV}$ estimate in the centre of the galaxy. For this reason we ran the fitting procedure with an alternative dust surface density map (see Fig.~\ref{fig:dustmaps}).

Since the standard model lacks dust emission in the center of the galaxy, we constructed a dust map that is strongly peaked in the center. To do so we normalize the PACS $70 \, \mu$m flux density map, which is a good tracer of the total infrared emission \citep{Galametz2013}. At this wavelength, we are more sensitive to warmer dust, but we importantly attain sufficient spatial resolution ($\sim 300$ pc) in our model. If we were to derive a dust mass map based on the SPIRE bands or based on an modified black body fit to the \textit{Herschel} data, the effective resolution would quickly drop to 1 kpc. The PACS $70 \, \mu$m  based dust map is significantly different from the $A_{FUV}$ based dust map, which at least allows a qualitative assessment between both models.

\begin{figure*}
	\includegraphics[width=0.97\textwidth]{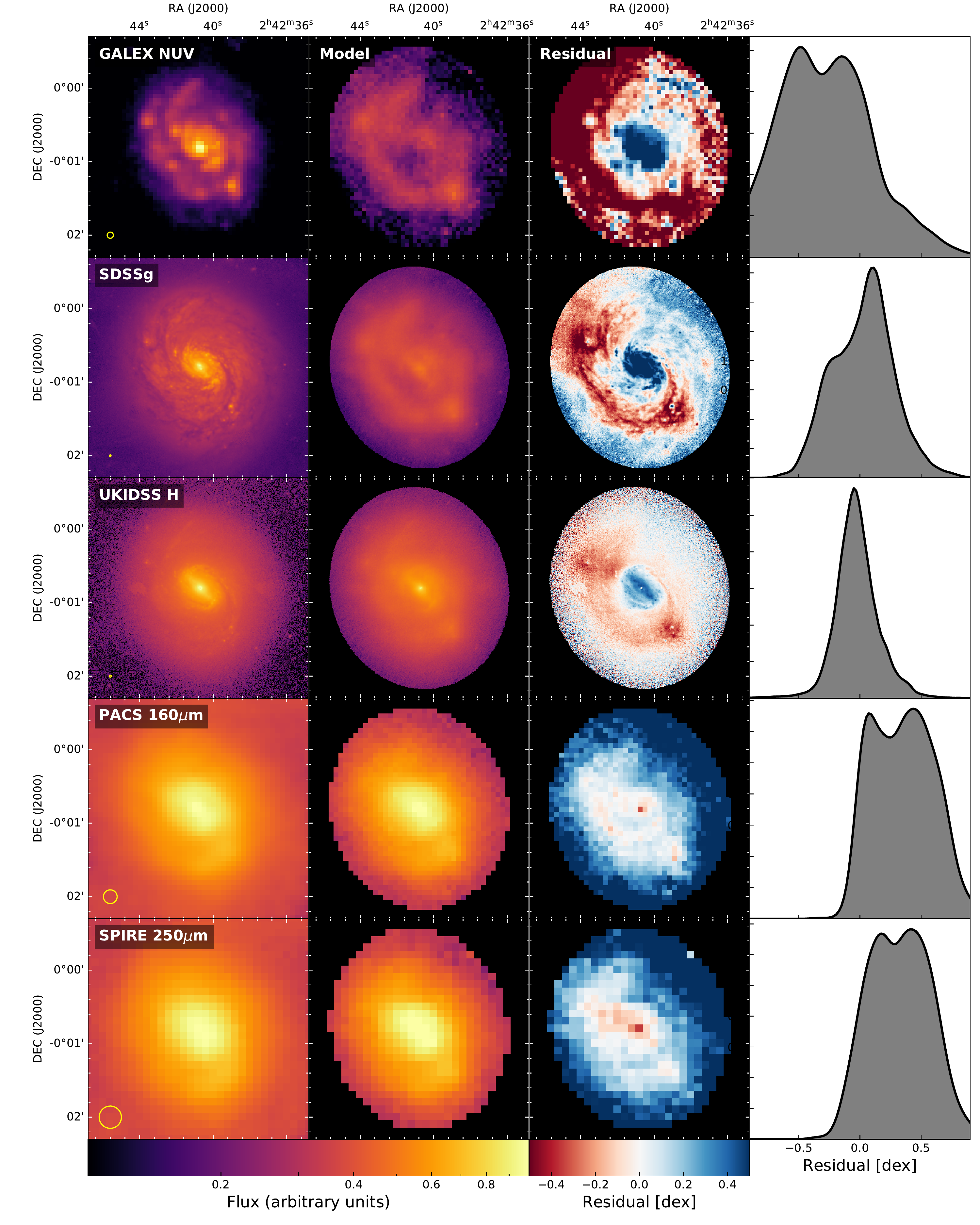}
    \caption{Same as Fig.~\ref{fig:residuals} but for the alternative dust mass map. Spatial comparison of the model to observed broad-band images (left column). The second column contains the model images. Note that the model images are not convolved by any PSF and only hold the intrinsic model PSF (a composite of the input images). Residual images are shown in the third column. Corresponding residual distributions are represented by KDE plots in the right column. The area under the KDE curves is normalized to 1.}
    \label{fig:residuals_alt}
\end{figure*}

In Fig.~\ref{fig:residuals_alt} we compare the observed and model images of the alternative model. This figure can be compared to Fig.~\ref{fig:residuals} for the standard model. We see a roughly opposite trend compared to the standard model. Instead of a central dip in the dust emission, the emission now peaks in the center, which was the goal of this new model. However, the central peak is too strong and overestimates the FIR observations. Consequently, the attenuation in the UV and optical bands is severe and causes a strong underestimation of the flux in the central regions. At the same time the attenuation in the spiral arms is not strong enough causing a large negative residuals there. The KDE distributions shown in the right column of Fig.~\ref{fig:residuals_alt} as a result show broad and often double-peaked residuals. We conclude that the alternative model is highly unbalanced and not suitable for further analysis.

\end{document}